\documentclass[12pt,a4paper,fleqn]{article}

\usepackage[textheight=23cm,textwidth=15cm]{geometry}
\usepackage{amsmath, amssymb, longtable, multicol, multirow, supertabular, afterpage} 
\usepackage{caption, subcaption}
\usepackage{graphicx}
\usepackage[table]{xcolor} 
\usepackage{makecell} 
\usepackage{url}
\usepackage[american]{babel}
\usepackage{here}
\usepackage[articletitle=true, super=false, chaptertitle=true]{achemso}
\setcitestyle{square}

\begin{document}

\begin{center}

{\LARGE\bf
 Chemoton 2.0: Autonomous Exploration of Chemical Reaction Networks
}

\vspace{1cm}

{\large
Jan P. Unsleber\footnote{ORCID: 0000-0003-3465-5788},
Stephanie A. Grimmel\footnote{ORCID: 0000-0001-7633-6123}, and
Markus Reiher\footnote{Corresponding author; e-mail: markus.reiher@phys.chem.ethz.ch; ORCID: 0000-0002-9508-1565}
}\\[4ex]

Laboratorium f\"ur Physikalische Chemie, ETH Z\"urich, \\
Vladimir-Prelog-Weg 2, 8093 Z\"urich, Switzerland

February 25, 2021

\vspace{.43cm}

\textbf{Abstract}
\end{center}
\vspace*{-.41cm}
{\small
Fueled by advances in hardware and algorithm design, large-scale automated explorations of chemical reaction space have become possible.
Here, we present our approach to an open-source, extensible framework for explorations of chemical reaction mechanisms based on the first principles of quantum mechanics. 
It is intended to facilitate reaction network explorations for diverse chemical problems with a wide range of goals such as mechanism elucidation, reaction path optimization, retrosynthetic path validation, reagent design, and microkinetic modeling.
The stringent first-principles basis of all algorithms in our framework is key for the
general applicability that avoids any restrictions to specific chemical systems.
Such an agile framework requires multiple specialized software components of which
we present three modules in this work.
The key module, \textsc{Chemoton},drives the exploration of reaction networks.
For the exploration itself, we introduce two new algorithms for elementary-step searches that are based on Newton trajectories.
The performance of these algorithms is assessed for a variety of reactions characterized by a broad chemical diversity in terms of
bonding patterns and chemical elements. We reproduce and significantly extend what is known about these reactions and provide the resulting data to be used as a starting point for further explorations and for future reference.
}

\newpage

\section{Introduction}
\label{sec:introduction}

Chemical reaction mechanisms are at the core of our understanding of chemical processes.\cite{Fialkowski2005}
At the level of elementary reaction steps, reaction mechanisms quickly expand into
large reaction networks. Whereas a complete overview of all relevant elementary
reactions is mandatory for a correct kinetic picture of a chemical process,
such an overview is seldom available. This lack of detail is rooted in limited experimental information available on a specific process and in limited
quantum chemical data that is difficult to gather by manual exploration. 
To alleviate this problem, autonomous exploration algorithms that rest on fully
automated procedures have been developed in the past decade.\cite{Sameera2016,Dewyer2017,Vazquez2018,Simm2019,Unsleber2020,Maeda2021}

Individual elementary reaction steps can be obtained by well-established 
quantum chemical approaches\cite{Cramer2006,Jensen2006}, but such calculations
require a significant amount of 
manual work. To map out thousands of such steps is therefore not feasible. Moreover,
it is also not sensible to spend human effort on tasks that can be
largely automatized. It is therefore highly desirable to establish
frameworks for the autonomous construction of reaction networks by automated 
identification of a huge number of 
potentially important elementary steps that requires as little human interference
as possible. 
Here, we present such a framework as a free open-source software which rigorously
relies on first-principles quantum chemical methods and concepts in such a way that
its range of applicability to chemical systems is, in principle, not limited.

This work is organized as follows: In the next section, we present the layout of our software framework designed for the exploration of reaction networks.
We then detail the algorithms and procedures implemented in its modules and how they are applied to elucidate reaction pathways in an automated way.
After a description of the computational methodology, we consider specific examples of chemical processes that are well known in the literature to analyze and highlight the specific features of our framework for autonomous reaction network exploration.

\section{Software}
\label{sec:theory}

\subsection{Background}
During the last years, we have been developing the software environment 
\textsc{SCINE}\cite{Scine} ("Software for Chemical Interaction Networks")
for quantum chemical calculations
with a special focus on algorithmic stability, automation, interactivity, efficiency,
and error control.\cite{Haag2014,Vaucher2016,Muhlbach2016,Vaucher2017,Husch2018a,Husch2018,Vaucher2018,Heuer2018,Simm2018}
One of its capabilities has been the autonomous exploration of reaction networks based
on first principles\cite{Bergeler2015}, which was realized in the \textsc{Chemoton} module\cite{Simm2017}.
Since establishing the \textsc{SCINE} environment required the development
of new algorithms and concepts, their interoperability was difficult to guarantee and to maintain.
It is for this reason, that a complete re-design of the \textsc{Chemoton} and related modules was necessary
in order to honor the existing developments and to provide a safe harbor for new ones.
\textsc{Chemoton} 2.0 now strives to deal with key challenges of reaction exploration automation:\cite{Unsleber2020}
(i) the autonomous operation on huge sets of raw reaction data, (ii) minimal expectations on the operator side regarding the technical details of explorations, (iii) generally unknown degrees of completeness and (iv) accuracy of the explored data.
In this work, we describe this new layout and elaborate on its
specific features and new algorithms.

Note that many groups have worked on the development of such type of exploration software;
examples are
\textsc{NetGen}\cite{Broadbelt1994},
\textsc{GRMM}\cite{Maeda2013},
\textsc{ZStruct}\cite{Zimmerman2013a},
\textsc{RMG}\cite{Gao2016,Liu2021},
\textsc{AARON}\cite{Guan2018},
\textsc{KinBot}\cite{VandeVijver2020},
\textsc{ChemTrayZer}\cite{Krep2022},
\textsc{Nanoreactor}\cite{Wang2014},
\textsc{autodE}\cite{Young2021},
\textsc{YARP}\cite{Zhao2021},
\textsc{AutoMeKin2021}\cite{Martinez-Nunez2021},
and many more\cite{Rappoport2014,Kim2014,Suleimanov2015,Habershon2016,Kim2018,Grambow2018,Grimme2019,Rizzi2019,Jara-Toro2020,Schmitz2021,Chen2022}.

In the light of these developments, we emphasize that our \textsc{Chemoton} project
aims at establishing a software framework that is out-of-the-box applicable to any kind of chemical
problem and can host any algorithm that was shown to be reliable and efficient.

\subsection{General Software Layout}
Each SCINE module features a specific set of tasks with clean interfaces facilitating rapid prototyping and long-term maintainability.
There are three modules that we present in this work.
These three modules are:
(i), the front-end, which comprises all parts of the exploration software that an operator will interact with regularly during exploration tasks.
(ii), the backend, which is the part of the software that carries out all calculations.
(iii), the data storage.

\begin{figure}[htbp]
 \begin{center}
  \includegraphics[scale=.50]{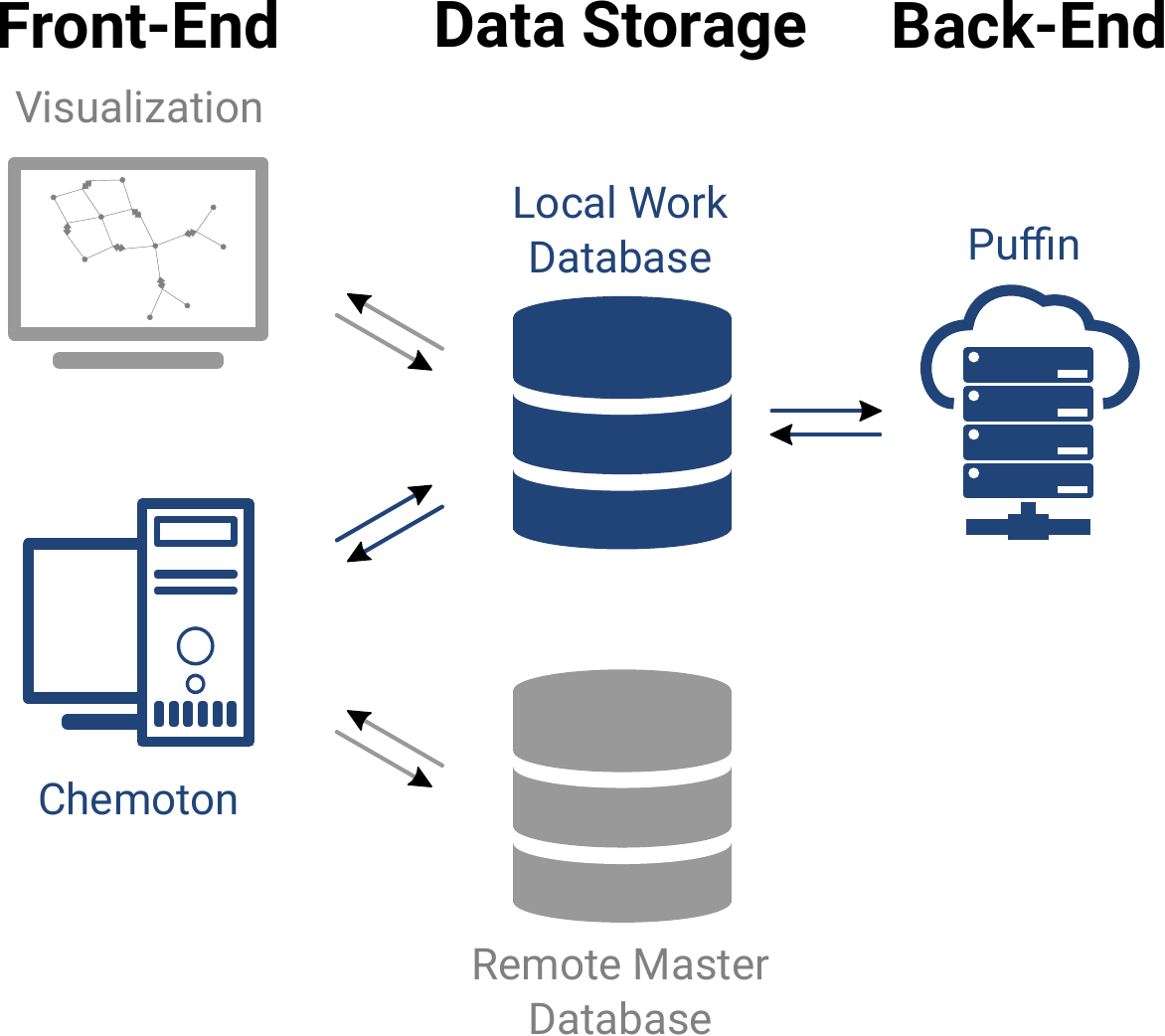}
 \end{center}
 \caption{\label{fig:scine_schematic}\small Schematic overview of software and information flow in the \textsc{SCINE} framework, showing \textsc{Chemoton} as the key steering element in the exploration. Gray parts represent software that is currently under development.}
\end{figure}

\begin{figure}[htbp]
 \begin{center}
  \includegraphics[scale=.30]{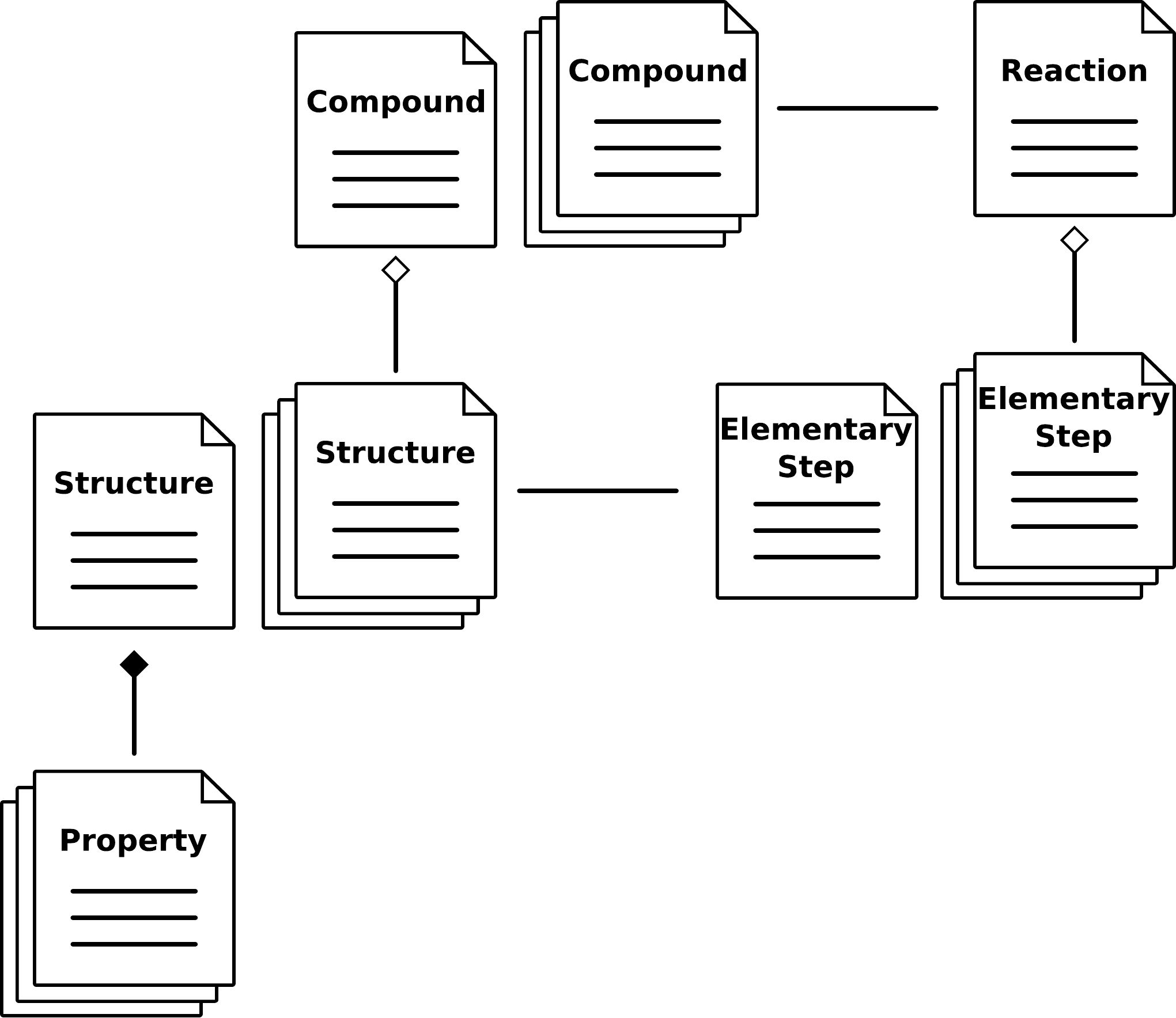}
 \end{center}
 \caption{\label{fig:db_layout_1}\small Scheme of aggregation (hollow connector), composition (filled connector) and association (line) in the data generated and stored as properties, structures, elementary steps, compounds, and reactions in the \textsc{Scine} database.}
\end{figure}

As can be seen in Figure~\ref{fig:scine_schematic}, there is a distinct flow of data between parts in the aforementioned stages, a local database being the central facilitator of the data flow.
The layout of the database is modeled after the concepts discussed in Ref.~\citenum{Unsleber2020} and in the  remainder of this work we shall refer to the definitions of
\textit{structure}, \textit{compound}, \textit{elementary step}, and \textit{reaction} 
given therein. To emphasize that we specifically refer to these definitions and to distinguish them from their colloquial use, we print the technical terms according to Ref.~\citenum{Unsleber2020} in an italic font.
\textit{Structures} are defined as a point on a potential energy surface, with fixed atom count, atomic positions, number of electrons, and spin.
\textit{Compounds} are a group of \textit{structures} with the same atom counts, connectivity (bonds), charge, and spin.
An \textit{elementary step} is defined as a rearrangement of bonds or transfer of electrons that connects an educt valley with a product valley through a single transition state.
\textit{Reactions} are groups of \textit{elementary steps}, all connecting different \textit{structures} of the same \textit{compounds} reacting with each other.
Hence, there are \textit{structures} that are aggregated into \textit{compounds}, and \textit{elementary steps} that are aggregated into \textit{reactions}.
\textit{Structures} are connected via \textit{elementary steps} and \textit{compounds} are connected via \textit{reactions}.
Hence, \textit{elementary steps} and \textit{reactions} are associated with \textit{structures} and \textit{compounds}, respectively.
This data structure is depicted in Figure~\ref{fig:db_layout_1}.

\begin{figure}[htbp]
 \begin{center}
  \includegraphics[scale=.30]{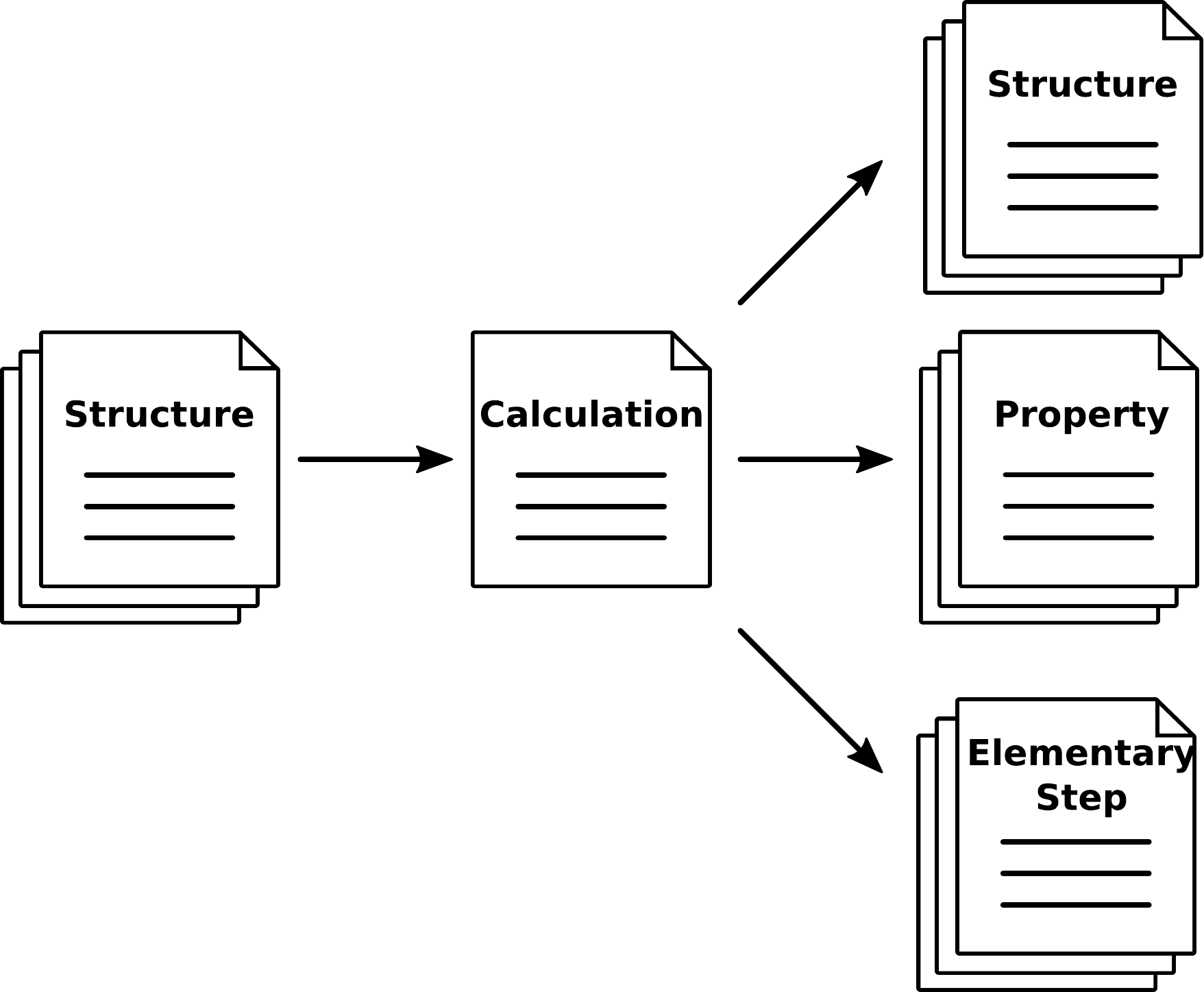}
 \end{center}
 \caption{\label{fig:db_layout_2}\small Schematic representation of data usage and generation based on calculations present in the database.}
\end{figure}

Additionally, there are \textit{calculations} and \textit{properties} stored within the database. 
As can be seen in Figure~\ref{fig:db_layout_1}, \textit{structures} have \textit{properties} and \textit{properties} therefore only exist in reference to a \textit{structure} (composition).
\textit{Calculations} are orders to be executed in the backend.
Based on a given set of \textit{structures}, they then generate new \textit{properties}, \textit{structures}, and \textit{elementary steps}.
This workflow is depicted in Figure~\ref{fig:db_layout_2}.\\
\textit{Calculations} can be simple quantum chemical energy evaluations of a fixed molecular structure, or a more involved chain of tasks.
For example, one may request a molecular dynamics simulation with snapshot extraction and thermodynamic integration as a single \textit{calculation}.\\
Similarly, \textit{properties} can be a simple scalar, such as an electronic energy, or they can be higher dimensional data, such as a density matrix, a bond order matrix, or the normal modes of a nuclear Hessian.

To distinguish the different (quantum chemical) methods generating data for chemical systems, all structures and properties are tagged with a \textit{model} with which they were generated with.
All calculations and subsequently all properties are tagged with the a database representation of that \textit{model} allowing for easy distinction and comparison.
The number of possible settings and finer details of \textit{calculations} are generally large and can vary significantly, depending on the actual type of \textit{calculation} and its implementation in the backend.
For this reason the \textit{model} only tracks high-level information, and an additional set of finer settings is used inside each \textit{calculation}.

Although this high-level/low-level split of information appears arbitrary at first sight, the idea is to include only key information in the \textit{model} data structure so that properties generated with an identical \textit{model}, but possibly with different settings in their \textit{calculation}s, can still be expected to be comparable.
For example, a DFT functional will be listed in the model because a switch of functional would make data across two calculations non-comparable.
Slightly tighter convergence criteria or a reasonably modified integration grid size will generally still result in comparable data across runs and thus not be listed in the model, but only in the \textit{calculation}'s settings.

Each unique \textit{structure} is identifiable by a unique database identifier (DB-ID), as is any other document stored in the database. 
However, much of chemical interpretation and understanding is based on the differences of these structures, \textit{i.e.}, of their three-dimensional arrangement of atoms and also their bond patterns.
To allow for better and more efficient comparisons that do not require direct comparisons of the stored three-dimensional structure, all \textit{structure} objects can be tagged with additional strings.
This dictionary of tags is meant to contain simpler, possibly non-unique identifiers, mostly graph representations.
SMILES\cite{Krenn2020}, SELFIES\cite{Weininger1988}, and InChI\cite{Inchi2001}, but also a IUPAC name may be added here.
Internally, the \textsc{Chemoton} framework relies heavily on a specialized, serialized string representation of the molecular graph as it is generated by the SCINE module \textsc{Molassembler}\cite{Sobez2020, Sobez2020a}.
The particular graph representation generated has the advantage that permutations in the list of atoms do not affect it, and hence, graph comparisons can be carried out on the database side as simple string comparisons.

A straightforward alternative to this approach would be an expensive comparison of root mean square deviations (RMSD) of nuclear coordinates.
One short-coming of the RMSD is its dependency on the size of the underlying molecules.
Furthermore, these comparisons would have to be made invariant to atom permutations, which would make them even more computationally demanding.
As a result, such RMSD-based comparisons would have to be done outside the database framework, and hence, they become unfeasible for massive amounts of data.

By contrast, a \textsc{Molassembler}-based graph (plus charge and multiplicity) allows us to efficiently assess whether two structures are part of the same compound within a database-side query.
Furthermore, additional information is stored in the same list of representations, allowing for the determination and comparison of different conformers of the same \textit{structure}.

Unfortunately, the \textsc{Molassembler} graph is not readable by humans in its serialized form.
However, for our purposes here, readability of the graph string is of low priority, because the key purpose of our set-up is to efficiently sort and deduplicate \textit{structures}.
Furthermore, the graph can be translated into a vector graphic displaying the molecular structure with a very simple \textsc{Python3} script (see Ref.~\citenum{Sobez2020} and Ref.~\citenum{Sobez2020a} for details).

\begin{figure}[htbp]
 \begin{center}
  \includegraphics[width=1.0\textwidth]{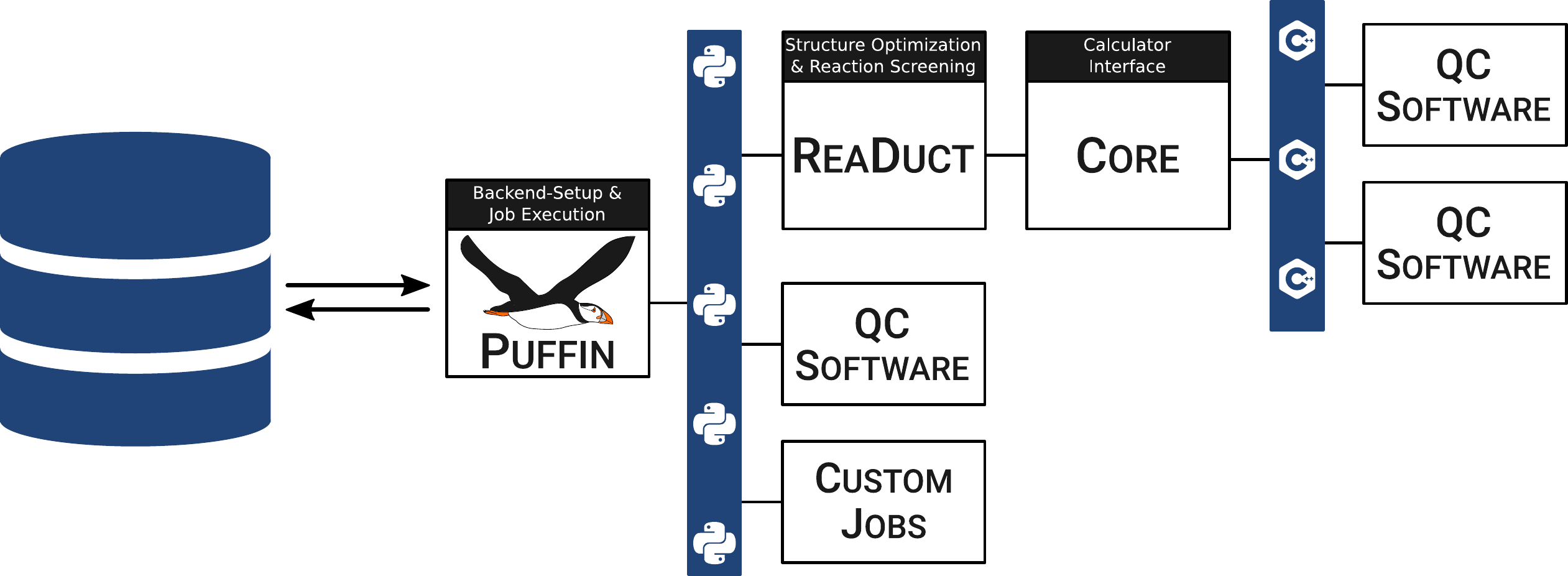}
 \end{center}
 \caption{\label{fig:backend}\small Interface layers in the backend defined by the \textsc{Python3} software package \textsc{Puffin}.}
\end{figure}

The backend itself has a multi-layered design, which we briefly outline here; see also Figure~\ref{fig:backend}.
At the center of the backend is a \textsc{Python3} package, which we call \textsc{Puffin}.  \textsc{Puffin} directly interacts with the database and defines all executable tasks that can possibly be performed.
In this work, the words 'job', 'task', and 'calculation' all denote a computation carried out in the backend by a \textsc{Puffin} instance.

\textsc{Puffin} maintains a list of jobs that can be extended by writing \textsc{Python3} source code. It therefore defines an abstraction layer for high level tasks in \textsc{Python3}.
The jobs can vary in terms of their complexity, and also the general style of linkage between programs is not restrictive (\textit{e.g.}, compiled and packaged linkage as well as scripting for installed programs in a local environment are available).

Many raw-data production jobs will rely on quantum chemical calculations to be carried out with standard quantum chemistry methods.
In order to abstract from the choice of a specific quantum chemistry program, the \textsc{SCINE Core} C++ package defines an interface for a generic quantum chemistry calculator.
The \textsc{SCINE Utilities}\cite{Bosia2020} package provides utilities to facilitate the quick implementation of an instance of the interface.
The required features of these calculators (instances) are based on single-point calculations for a given set of nuclear coordinates.
A calculator only needs to provide basic features, such as generating an energy, partial charges, nuclear gradients, and possibly a nuclear Hessian and bond orders, to be a fully defined instance of the interface.
Currently, \textsc{xTB}\cite{Bannwarth2020}, \textsc{Sparrow}\cite{Husch2018, Bosia2020b}, \textsc{Serenity}\cite{Unsleber2018, Barton2020}, \textsc{ORCA}\cite{Neese2012, Neese2018}, \textsc{Gaussian}\cite{Frisch2016}, and \textsc{Turbomole}\cite{Turbomole2019} are available via these interfaces.

For all of these calculators, the \textsc{ReaDuct} program\cite{Vaucher2018, Brunken2020} provides the higher-level
implementations of structure manipulations and optimizations such as local energy minimum localization, transition state search, and so forth.
Puffin then defines a default pipeline for the \textsc{Chemoton}-specific tasks using \textsc{ReaDuct} which interfaces to the quantum chemistry program packages for raw data generation.
With these two layers of abstraction it is possible to allow for a great variety of raw-data production programs to be activated in an exploration workflow.

For good computational performance, each \textsc{Puffin} instance allocates a fixed amount of cores, memory, and disk space, which 
can then parallelize the underlying calculations of a requested job.
However, the present version does not support single \textsc{Puffin} instances parallelized across nodes.

Assuming a large number of calculations that have to be processed when exploring reaction networks, it is important to tackle parallelization first  by exploiting the inherent trivial parallelism and to simply launch many calculations simultaneously, rather than parallelizing the individual calculations (which, however, may be done through the raw data generating quantum chemistry packages).
To facilitate the trivial parallelism, \textsc{Puffin} instances are designed to be packaged in containers. 
As an example, most of the data shown in the later sections has been generated in a high performance computing environment running \textsc{Singularity}\cite{Singularity} images scaled up to hundreds of \textsc{Puffin} instances at the same time.
Other containerization software such as \textsc{Docker}\cite{Merkel2014} or \textsc{Podman}\cite{Podman} are also supported.

Given that not all of the interfaced quantum chemical software is available free of charge and/or open-source, we only provide images or files creating containers for those programs that are open source and document their extension for closed source quantum chemistry software.
For more details, we refer the reader to the documentations of the \textsc{Puffin} \textsc{Python3} package.\cite{puffinZenodo}

\subsection{Chemoton 2.0}
We now describe version 2.0 of \textsc{Chemoton}, which is the software module that drives the machinery for autonomous chemical reaction network exploration.
\textsc{Chemoton} defines two major objects that drive the exploration; these we dubbed engines and gears.
Engines model potentially infinite loops of the same action.
For example, an engine may be tasked with the generation of conformers.
It then continuously updates new compounds with all conformers that are part of this compound.

Gears are the specific algorithms that are engaged by each engine.
Continuing with the conformer-generation example, it is possible either to run molecular dynamics simulations of a given structure to extract all realized conformers by clustering or to generate conformer guesses by explicit construction and enumeration. Then, these propositions for conformers can be optimized and deduplicated to generate the final set of conformers.
These two options constitute two gears that could be attached to the engine for managing the conformer generation.

A key feature of any software that explores chemical reaction networks must be an option to trim the combinatorial explosion of potentially reactive events that must be inspected when exploring the network.
Accordingly, there are engines in \textsc{Chemoton} that allow for filters to be applied in such a way that they only process a part of the proposed trials in search for successful reactive events.
As an example, consider a simple filter that implements an upper limit to the molecular weight of structures that are considered for the exploration.

This one and other filters can be combined with the common logical operators 'and' and 'or', resulting in fine-grained control of the exploration process.
The filters can be extended on the fly by the operator in such a way that they allow for easy tailoring to the specific reactive system under consideration.

The set of filters is under constant development. Examples to be considered are filters that are based on semi-local information and/or kinetically modelled data of given compounds. For instance, whereas the general setting of \textsc{Chemoton}  considers all structures in the network (existing and emerging ones) as potential reactants, this explosion of options for successful reactive events may be tamed by noting that compounds will need to be sufficiently long-lived to undergo a reaction with some other reactant. Hence, if some stable intermediate structure is surrounded by barriers that are easily overcome under reaction conditions, its reactivity with other structures further apart in the network (irrespective of whether they are stable or fleeting ones) does not need to be probed. In fact, the simple network exploration without proper kinetic modeling that includes diffusion and mass transport cannot properly assess the importance of such reactions, and it can, therefore, be advantageous to not allow for them in the network exploration.
We intend to elaborate on these features in future releases so that a range of techniques for steering an exploration will be available.

The actual driving of \textsc{Chemoton} to facilitate an exploration goal then boils down to the combination of engines, gears, and filters that are driving and guiding the exploration.

\begin{figure}[htbp]
 \begin{center}
  \includegraphics[width=1.0\textwidth]{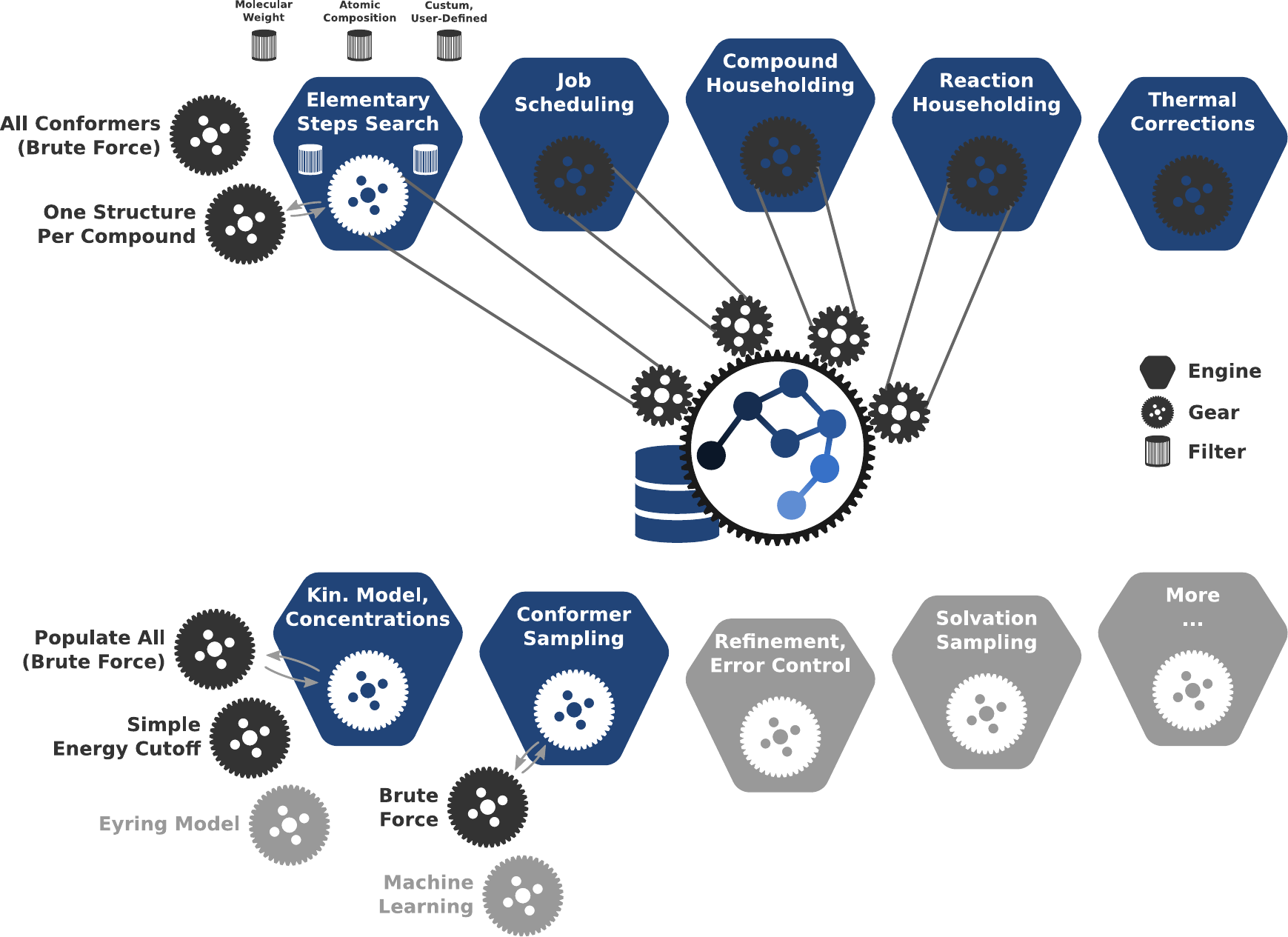}
 \end{center}
 \caption{\label{fig:chemoton}\small Schematic structure of \textsc{Chemoton}. Gray components are not available in the version 2.0 release but are currently under development.}
\end{figure}

To explain the engine, gear, and filter structure in more detail, we consider the example of a basic forward exploration starting from two known compounds defined on input (the
overall structure of this set-up is shown in Figure~\ref{fig:chemoton}):
Assuming two molecular structures were provided in the database for the start of an exploration, the first engine addresses compound book keeping.
Starting this engine will optimize, sort, and deduplicate any new structures (user given or discovered) into compounds.

For the backend to run requested calculations we enable the job scheduling engine.
Any optimization or graph determination job will be scheduled and will only be run by the backend if this engine allows them to be executed.
This engine enables prioritization of job classes (\textit{e.g.}, all geometry optimizations are preferred over Hessian evaluations) and can delay or limit job executions if the operator asks for this.
In the generic case, the engine will schedule all requested jobs as they are generated.

As soon as there are the first compounds available, an engine for elementary step searches can start generating elementary step trials.
This engine in its simplest form may use a gear that only considers one conformer per compound and allows for all compounds to be considered in elementary steps trials;
it will then consider both inter- and intramolecular reactions of the given starting materials. In this way, one avoids a combinatorial explosion of many conformers considered as reactants right from the
start as this would choke the exploration. However, conformational resolution can be activated at a later stage in the exploration process.
In a more advanced setup, filters may be applied to limit the compounds that are used (\textit{e.g.}, by molecular mass) or to limit the atoms that are considered as reactive (\textit{e.g.}, by local symmetry).

To complete the minimal setup, a book keeping engine for reactions will be added that sorts elementary step into existing reactions or creates new ones if none are matching.
With this setup, intramolecular  as well as intermolecular reactions of the starting compounds ($A\rightarrow ~?$) are probed ($A+A\rightarrow ~?$, $A+B\rightarrow ~?$).
Except for one alteration, this setup is the one that was used multiple times to generate the data for the results section of this work.
The alteration is that depending on the expected reaction outcome either inter- or intramolecular reactions were probed, not both.

If it is intended to feed any newly found compounds back into the list of possible compounds that can react, an engine that evaluates the reaction kinetics and deems compounds accessible must be enabled.
A default engine that simply enables all compounds and one that employs a simple energy cut-off criterion for reaction barriers are available for \textsc{Chemoton} 2.0.

With this work, we release \textsc{Chemoton 2.0} with a feature set that allows one to run most exploratory tasks in a basic fashion.
In future releases, microkinetic modeling of the growing network will be made available on the fly by combination with our \textsc{KinNetX} module\cite{Proppe2019}.
Additionally, there are engines/gears that may automatically generate conformer ensembles for all compounds and complete thermodynamical data within standard models of statistical thermochemistry by Hessian calculations so that
Gibbs free energies can be computed for all stationary points of the reaction network.
Engines/gears that refine existing data points with advanced electronic-structure models\cite{Simm2018} and that sample explicit solvents in our subsystem microsolvation approach \cite{Bensberg2022} will also be made available in future releases.

\textsc{Chemoton} as the driver of explorations is a \textsc{Python3} package and thus a text-based program that requires scripting. Currently, this may be taken as a major hurdle in the familiarization phase with the software.
However, a graphical user interface is currently under development in our group and this also will be made accessible open source in due time.
The graphical user interface will streamline the most common tasks and exploration types for non-expert users, visualize reaction networks, and monitor computational statistics.

Moreover, eventually even a remote database will be established and made available in order to allow for the central storage of high-accuracy, quality-validated, computationally expensive data.
The introduction of this remote master database will then require \textsc{Chemoton} to replace request for new calculations with imports of existing data.
Continuous uploads of relevant data will then greatly reduce computational cost of running explorations that involve ubiquitous chemical (sub)networks.
Clearly, this will reduce the exploration effort and time as well as energy consumption; as a byproduct it will generate a valuable repository for data driven approaches.

The current setup also allows for user intervention into the exploration process, which we consider necessary as full autonomy is in principle possible for an exploration process, but will face severe resource requirements in practice.
Given that all exploration steps are run in recurring loops, extensions, such as manually inserting reactive species and elementary steps, can be picked up seamlessly.
Real-time refinement and drop-in of user explored reactions by means of, \textit{e.g.}, a haptic device\cite{Marti2009,Haag2013,Haag2014,Simm2019} are currently under development.

\section{Finding New Reactions}\label{sec:trial_math}
In this section, the default exploration algorithm in \textsc{Chemoton} is described. 
This algorithm searches for elementary steps (elementary-step trials) starting from existing structures in the data base.
Found elementary steps are then assigned to reactions (either they define a new reaction or they are another path for a known reaction to occur).
Many single-ended and multi-ended time-independent search algorithms as well as molecular dynamics based approaches have been described in the literature that can be used for this purpose\cite{Broadbelt1994,Maeda2013,Zimmerman2013a, Gao2016, Liu2021, Guan2018, VandeVijver2020, Young2021,Zhao2021,Martinez-Nunez2021,Rappoport2014,Kim2014,Wang2014,Suleimanov2015,Habershon2016,Kim2018,Grambow2018,Grimme2019,Rizzi2019,Jara-Toro2020,Schmitz2021} and our implementation is general enough to accommodate many of them (up to the point where they could be directly compared within one software framework).

The algorithm implemented for this work is a single-ended transition state search algorithm dubbed 'Newton trajectory scan'. It is inspired by the work of Quapp and Bofill\cite{Quapp2020} and of Maeda \textit{et al.}\cite{Maeda2010, Maeda2011,Maeda2014, Maeda2018}. Our algorithm's key step also bears some similarity with the relaxed surface scan available in the ORCA program\cite{Neese2012, Neese2018}, according to its description in the ORCA manual. 
In addition, we implemented the artificial force induced reaction (AFIR) algorithm\cite{Maeda2010, Maeda2011}.
Approaches with similar concepts using reduced gradients in Newton trajectories can be found in the literature.\cite{Quapp1998, Anglada2001, Crehuet2002, Hirsch2004, Bofill2011}

The complete algorithm for an elementary-step trial, \textit{i.e.} for finding a new elementary step, consists of the following steps:
\begin{enumerate}
    \item Reactive complex generation
    \item Newton trajectory scan
    \item Transition state guess extraction
    \item Transition state optimization 
    \item Intrinsic reaction coordinate (IRC) calculation
    \item Molecular graph comparison(s)
    \item Product optimization(s)
    \item Data analysis and storage
\end{enumerate}
Note that step two, the Newton trajectory scan, is the key step. 

\subsection{Reactive Complex Generation}
The algorithm used to generate reactive complexes from two given structures is in essence an extended version of the algorithm described in Ref.~\citenum{Simm2017}.
The current version allows to filter atoms and atom pairs of any single or pair of structures to be combined to yield a trial reactive coordinate, which itself may be subjected to a filtering logic.
These filters can be extended on the fly and can be combined through logically operators, as mentioned before.
They are designed to be based on general properties of the reacting structurees, such as first-principles heuristics\cite{Bergeler2015, Simm2017, Grimmel2019, Grimmel2021}.
They may be extended by machine learning algorithm for offline and on-the-fly learning of reactive and unreactive pairings.

\begin{figure}[htbp]
 \begin{center}
  \includegraphics[width=0.5\textwidth]{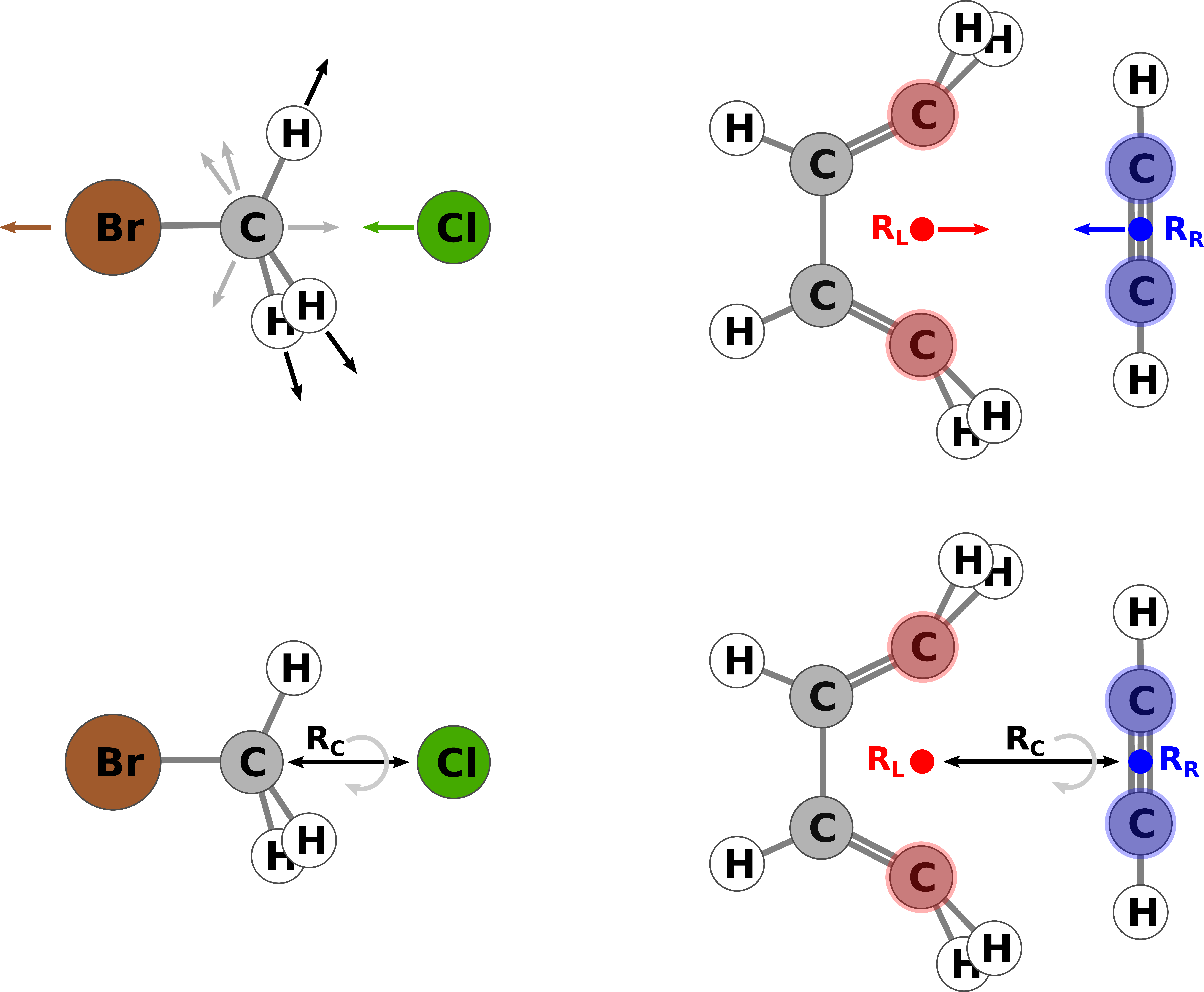}
 \end{center}
 \caption{\label{fig:rc_setup}\small Top row: Example of reactive sites for single atoms (left) and chosen atom pairs (right). Bottom row: Alignment with reaction coordinate $R_{\text{C}}$ and possible rotation for both examples.}
\end{figure}

For each atom or atom pair that survives the filters a reactive site is computed, as already described in Ref.~\citenum{Simm2017}.
Examples are shown in the top row of Figure~\ref{fig:rc_setup}.
The algorithm determining the reactive sites for each atom or atom pair is based mainly on steric hindrance and buried volume considerations.
It is a modified version of the algorithm described in Ref.~\citenum{Simm2017}.
For the details of the implementation we refer the interested reader to the source code released with this work.
The algorithm can be tuned to generate one or multiple reactive sites per atom or set of atoms.

The structures are then aligned along their reactive direction and it is possible to request any number of rotamers along the reactive coordinate.
Examples are shown in the bottom row of Figure~\ref{fig:rc_setup}.
While this holds for an intermolecular reaction with two reactants, for intramolecular reactions there is, obviously, no need to align reactants with respect to one another, and hence, the reactive complex structure is equivalent to the starting structure. Apart from that, reactive coordinates for intramolecular reactions, including dissociations, are set up analogously to intermolecular ones.

\subsection{Newton Trajectory Scans}\label{sec:nt}
In the next step, one or two molecular structures are forcibly distorted, in an attempt to generate a transition-state guess.
We implemented two variants of the particular procedure that we call Newton trajectories.
The two algorithms mainly differ in the way reactive atoms can be selected and are combined into one trial reactive coordinate.
For clarity, we will refer to all trial coordinates as (trial) reactive coordinates and to the final minimum energy path of an elementary step as reaction coordinate.

\subsubsection{Newton Trajectory Algorithm 1}
\label{sssec:NT1}
In the first Newton trajectory algorithm (NT1), starting with a reactive complex, two fragments (two sets of atoms, $\{L\}$ and $\{R\}$) are forced onto one another.
The two sets of reactive sites may contain any number of atoms.
In the simplest case, both sets consist of only a single atom each.
Then, the reactive coordinate $\mathbf{R}_C$ is the normalized vector of the direction that connects the two atoms. The movement along this vector is constrained for both atoms.
The movement constraint is enforced by alteration of the 'natural' (true) gradient in a steepest decent optimization procedure.
The steepest decent or gradient decent optimization procedure updates the atom positions ($\mathbf{R}$) by subtraction of the total gradient ($\mathbf{g}$), scaled with some factor $\alpha_\text{SD}$.
\begin{equation}
    \mathbf{R}^{n+1} = \mathbf{R}^{n} - \alpha_\text{SD} \cdot \mathbf{g}
\end{equation}
With our chosen modification, the optimization then becomes a scan along the chosen reactive coordinate with simultaneous optimization of all other degrees of freedom for which the gradient is not altered.

The 'natural' gradient along the chosen reactive coordinate $\mathbf{R}_C$ is neglected, and instead, an artificial force ($\mathbf{F}_L$, $\mathbf{F}_R$) is applied:
\begin{equation}\label{eq:nt1_grad_l}
    \mathbf{g}^{\text{NT1}}_I = \mathbf{F}_L \quad \forall \: I \in \{L\}, \quad \text{with} \quad \mathbf{F}_L = -0.5 \cdot \mathbf{\alpha}_{\text{NT1}} \cdot \mathbf{R}_C
\end{equation}
and
\begin{equation}\label{eq:nt1_grad_r}
    \mathbf{g}^{\text{NT1}}_I = \mathbf{F}_R \quad \forall \:  I \in \{R\}, \quad \text{with} \quad \mathbf{F}_R = +0.5 \cdot \mathbf{\alpha}_{\text{NT1}} \cdot \mathbf{R}_C ~.
\end{equation}
This results in the constrained atoms moving towards one another and the entire scan moving the system energetically uphill.
All other components of the 'natural' gradient are followed as they are.

In this scan, we are free to choose $\alpha_\text{SD}$ and the magnitude of the applied artificial force $\alpha_\text{NT1}$.
As a result, we can tailor how much the atoms in the reactive coordinate move in each step of the scan, and we can tailor by how much they move in relation to all non-constrained atoms.
We have chosen to apply a factor of $\alpha_\text{SD}=0.5$ in Eqs.~\ref{eq:nt1_grad_l} and \ref{eq:nt1_grad_r} such that $|\mathbf{R_C}| / (\alpha_\text{SD} \cdot \alpha_\text{NT1})$ is the number of modified steepest decent cycles required until the constrained atoms collide.
This facilitates runtime estimations and ensures that atom movements are small enough to ensure straightforward convergence of single energy and gradient evaluations.

In case of multiple atoms building one or both reactive sites, the geometric center of each site (left: $\mathbf{R}_L$, and right: $\mathbf{R}_R$) is calculated,
\begin{equation}\label{eq:geometric_center1}
 \mathbf{R}_{L} = \frac{1}{N_{L}} \sum_{I \in \{L\}} \mathbf{R}_I \quad,
\end{equation}
and
\begin{equation}\label{eq:geometric_center2}
 \mathbf{R}_{R} = \frac{1}{N_{R}} \sum_{I \in \{R\}} \mathbf{R}_I.
\end{equation}
The reactive coordinate ($\mathbf{R}_C$) is then the vector between these two centers (see Eq.~\ref{eq:reactive_coordinate}), or the center and the other atom,
\begin{equation}\label{eq:reactive_coordinate}
 \mathbf{R}_{C} = \frac{\mathbf{R}_{L}-\mathbf{R}_{R}}{|\mathbf{R}_{L}-\mathbf{R}_{R}|}.
\end{equation}

\begin{figure}[htbp]
 \begin{center}
  \includegraphics[width=0.5\textwidth]{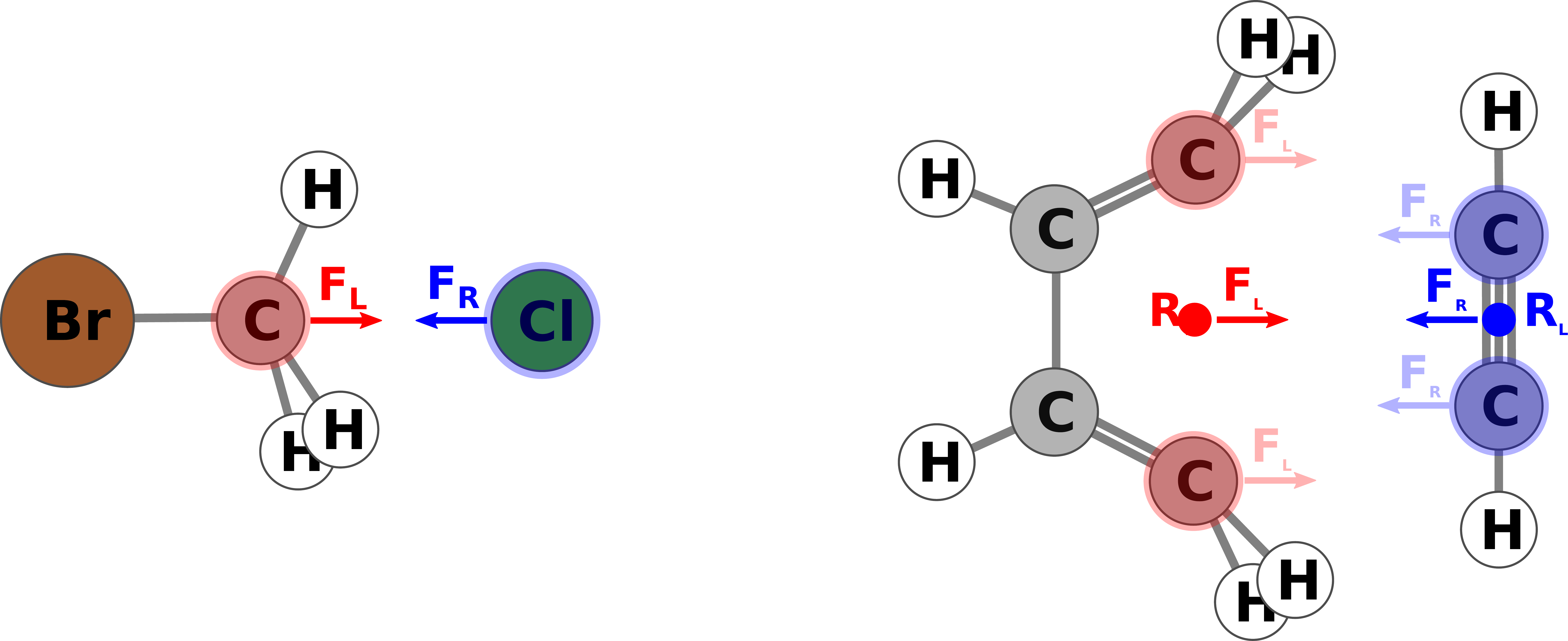}
 \end{center}
 \caption{\label{fig:nt_scan}\small Two example cases for a Newton trajectory scan with algorithm No.~1. Reactive atoms of the left-hand side are highlighted in red, those of the right-hand side in blue. Arrows indicate forces applied to the atoms.}
\end{figure}

In case of multiple reactive atoms in a reactive site, the gradient or force manipulation slightly changes.
All atoms in the reactive sites have their 'natural' gradients orthogonalized to the reactive coordinate, instead of it being set to zero.
This allows for movement perpendicular to the reactive coordinate, but constrains movement along the reactive coordinate.
All atoms in the reactive site are then subjected to the same artificial force ($\mathbf{F}_L$ or $\mathbf{F}_R$) as in the single-atom case, controlling the speed of collision.
The applied forces in these cases are calculated as
\begin{equation}
    \mathbf{g}^{\text{NT1}}_I = \frac{\mathbf{g}_I \cdot \mathbf{R}^T_C}{\mathbf{g} \cdot \mathbf{g}^T_I} \mathbf{R}_C + \mathbf{F}_L \quad \forall \:  I \in \{L\}, \quad \text{with} \quad \mathbf{F}_L = -0.5 \cdot \mathbf{\alpha}_{\text{NT1}} \cdot \mathbf{R}_C,
\end{equation}
and
\begin{equation}
    \mathbf{g}^{\text{NT1}}_I = \frac{\mathbf{g}_I \cdot \mathbf{R}^T_C}{\mathbf{g} \cdot \mathbf{g}^T_I} \mathbf{R}_C + \mathbf{F}_R \quad \forall \:  I \in \{R\}, \quad \text{with} \quad \mathbf{F}_R = +0.5 \cdot \mathbf{\alpha}_{\text{NT1}} \cdot \mathbf{R}_C,
\end{equation}
as indicated in Figure~\ref{fig:nt_scan}.

In order to allow for the unconstrained atoms to follow and adjust to the movement of the constrained atoms, microcycles of constrained geometry optimizations are run.
These microcycles are carried out with fixed reactive atoms and by default apply a standard BFGS\cite{Broyden1970, Fletcher1970, Shanno1970, Goldfarb1970} algorithm.

The scan will end if either of three conditions are met: i) a defined maximum number of gradient calculations is exceeded, ii) the constrained atoms are pushed into one another (by default, if any interatomic distance is smaller than $0.9$ times the sum of the tabulated covalent radii\cite{Cordero2008}) or iii) an SCF or gradient calculation fails.

Criterion iii) may seem like an obvious end without a result.
However, it should be noted that even if an SCF/gradient calculation does not converge in the later steps of the scan it may still be possible to extract a transition state from the previously generated trajectory.
In practice, any trajectory that has more than 5 steps is analyzed and it is tried to extract a transition state guess. 
The default number of steps of 5 is a result of the technical settings in the extraction algorithm, which is described below.
The entire first Newton trajectory algorithm as it is described here has been part of previous releases of \textsc{Readuct}\cite{Vaucher2018, Brunken2020}.

\subsubsection{Newton Trajectory Algorithm 2}
We have implemented another version of the Newton trajectory algorithm that, instead of defining two sets of reactive atoms that are pushed towards one another or pulled apart, sets of atom pairs are defined that are each pushed together or pulled apart.
In essence, this second algorithm, that we will also refer to as the NT2 algorithm, attempts to establish bonds directly where none exist or to dissociate existing bonds.
This clear focus on bonds may be easier to understand and, hence, to work with than defining an abstract group of atoms that are supposed to react.
Furthermore, the clear intent behind forming and dissociating certain bonds better lends itself to the generation and extraction of templates.
These advantages have been recognized and discussed in the literature, in particular in the work of Habershon and co-workers\cite{Ismail2019} 
which provided the inspiration for this second algorithm.

\begin{figure}[htbp]
 \begin{center}
  \includegraphics[width=0.5\textwidth]{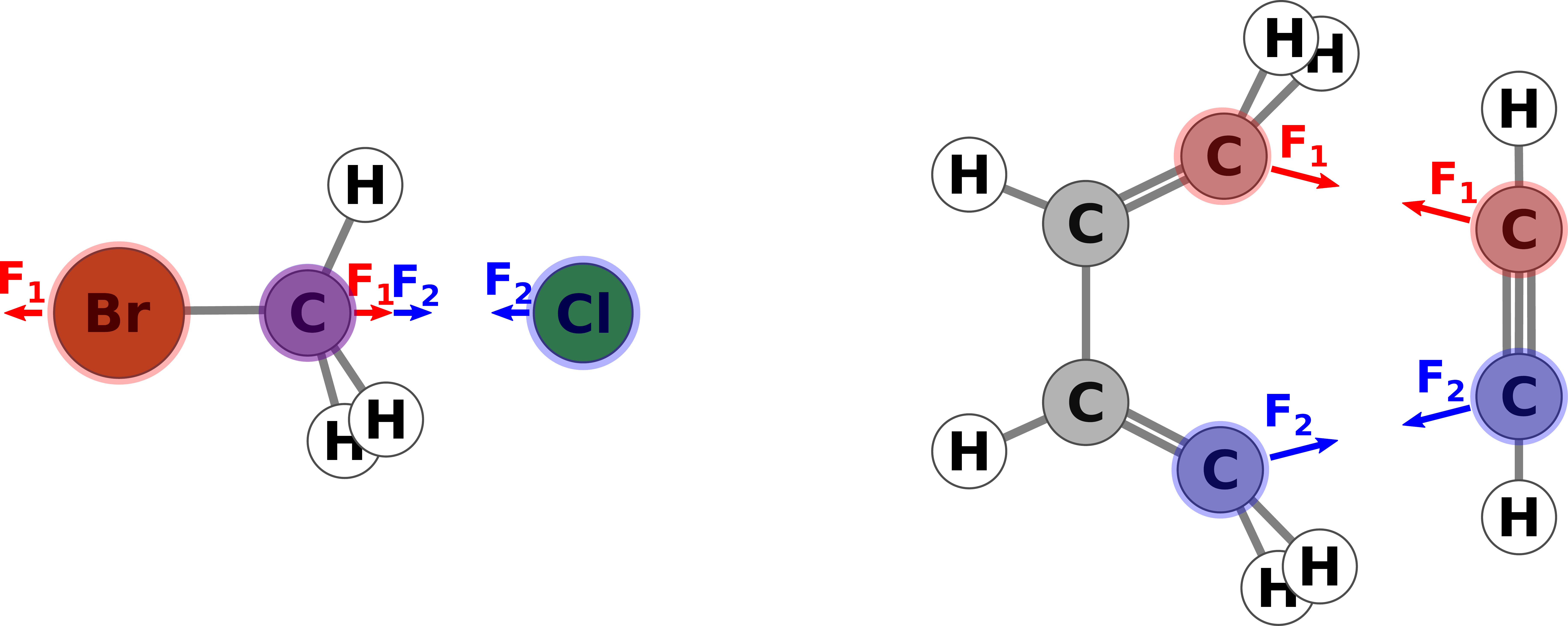}
 \end{center}
 \caption{\label{fig:nt2_scan}\small Two examples for a Newton trajectory scan with algorithm No.~2. Reactive-atom pairs are highlighted in red and blue. Arrows indicate forces applied to the atoms.}
\end{figure}

In the second Newton trajectory algorithm, atoms are constrained if they are part of any atom pair that is supposed to form a bond, or if they are part of an atom pair that is supposed to dissociate.
For all atoms for which either of these two conditions is fulfilled, the 'natural' 
atomic gradient contributions $\mathbf{g}_I$ are orthogonalized with respect to all constrained pair vectors $\mathbf{R}_{A,B}$ of which they are part of:
\begin{equation}
    \mathbf{g}_I \quad\rightarrow\quad \mathbf{g}^{\perp }_I, \quad \text{with} \quad \mathbf{g}^{\perp }_I \perp \mathbf{R}_{A,B} \quad \forall 
    \: \mathbf{R}_{A,B} \quad I \in \{A,B\}~.
\end{equation}

For cases with three or more constraints on one particular atom, this will likely result in $\mathbf{g}^{\perp }_I$ being a null vector.
For each constrained atom pair an additional force $\mathbf{F}_{A,B}$ is added to the orthogonalized atomic gradients $\mathbf{g}^{\perp }_I$, resulting in the final atomic gradients $\mathbf{g}^{\text{NT2}}_I$, 
\begin{equation}
    \mathbf{g}^{\text{NT2}}_I  = \mathbf{g}^{\perp }_I + 0.5 \cdot \frac{\alpha_{\text{NT2}}}{\max_{\forall A,B}(|\mathbf{F}_{A,B}|)}  \cdot \left( \sum_{B} \mathbf{F}_{I,B}
 - \sum_{A} \mathbf{F}_{A,I} \right)
\end{equation}
that is fed into the optimization procedure.
Again, microiterations, where all selected atoms are kept fixed, may be enabled between constrained optimization steps.

The forces added to the gradients are calculated as
\begin{equation}
    \mathbf{F}_{A,B} = \frac{|\mathbf{R}_{A,B}|-r^{\text{cov}}_A-r^{\text{cov}}_B}{|\mathbf{R}_{A,B}|} \mathbf{R}_{A,B}~,
\end{equation}
for bonds to be formed and
\begin{equation}
    \mathbf{F}_{A,B} = -\frac{1}{|\mathbf{R}_{A,B}|} \mathbf{R}_{A,B}~,
 \end{equation}
for bonds to be broken.
Here, $r^{\text{cov}}_I$ are the tabulated covalent radii\cite{Cordero2008} for the atom type of atom $I$.
The factor $\frac{|\mathbf{R}_{A,B}|-r^{\text{cov}}_A-r^{\text{cov}}_B}{|\mathbf{R}_{A,B}|}$ in the associative case is added to jointly move the requested atom pairs such that concerted reactions are the likely result, even if the initial interatomic distances of chosen pairs vary.

Similar to the first Newton trajectory algorithm, multiple stop criteria are defined.
If either of them is met, the resulting trajectory will be analyzed:
i) a defined maximum number of gradient calculations is exceeded,
ii) all atom pairs that were intended to form a bond have actually formed a bond (bond order $>0.75$) or they are crashing into one another (distance smaller than sum of covalent radii\cite{Cordero2008}), and all bonds intended to be broken are broken (bond order $<0.15$),
or iii) a SCF or gradient calculation fails.

The key advantage of this second algorithm is that it allows for a planned, simultaneous bond formation and dissociation, while the first algorithm may only generate these result serendipitously. As such, the algorithm is required to discover complex reaction pathways such as intramolecular rearrangement reactions on purpose.

A minor downside of this algorithm is, of course, the fact that the required definition of bonds relies on the dynamic calculation of bond orders, meaning that this algorithm requires one more property to be available from the chosen electronic structure method.
Furthermore, it is known that bond orders as commonly derived from standard electronic structure methods have issues\cite{Manz2017} and sensitive to the size of the chosen one-electron basis set\cite{Bridgeman2001}.

\subsection{Transition State Guess Extraction}
In order to extract a transition state guess from the trajectory generated by the scan, the electronic energy of all macroiterations is analyzed.
The data is passed several times through a 5-point Savitzky–Golay filter\cite{Savitzky1964}.
This eliminates small energy fluctuations and oscillations that might remain in the trajectory despite the BFGS microiterations.
Afterwards, the smoothed graph is analyzed and the structure corresponding to the highest energy maximum is extracted as the transition state guess for the next step.

\subsection{Finalization: The Remaining Steps}
The remaining steps are mostly straightforward applications of standard quantum chemical workflows.
First, a transition state optimization with a standard algorithm is carried out. By default, \textsc{Chemoton} employs a version of the algorithm proposed by Bofill\cite{Bofill1994}. 
Alternatively, an explicit eigenvector following algorithm\cite{Cerjan1981,Simons1983,Banerjee1985} or the dimer algorithm\cite{Henkelman1999,Heyden2005,Kaestner2008} can be chosen.

The resulting optimized transition state is validated by running an intrinsic reaction coordinate (IRC) scan and comparing the resulting molecular graphs for the forward and backward reaction with the initial structures in order to i) probe whether new products are formed and ii) ensure that the initial state is identical to the starting structures.
The graphs exploited are again the serialized \textsc{Molassembler} graphs employed throughout \textsc{Chemoton}.

Afterwards, the newly formed product molecules are separated (if there are more than one) and then optimized independently, in such a way that local minimum structures on their separated potential energy surfaces are obtained.
The separation of multiple new products is based on the graph analysis, with the attribution of electrons based on atomic partial charges and, by default, the assumption of a minimal total electron spin.
Finally, all results for the new products and the transition state are stored in the database.
Additionally, trajectory data of the elementary step is interpolated with a spline and stored in the database
(this step is motivated and discussed in the next section).

\subsection{Complexation: The Concept of Virtual Flasks}\label{sec:flask}

\begin{figure}[htbp]
 \begin{center}
  \includegraphics[width=1.0\textwidth]{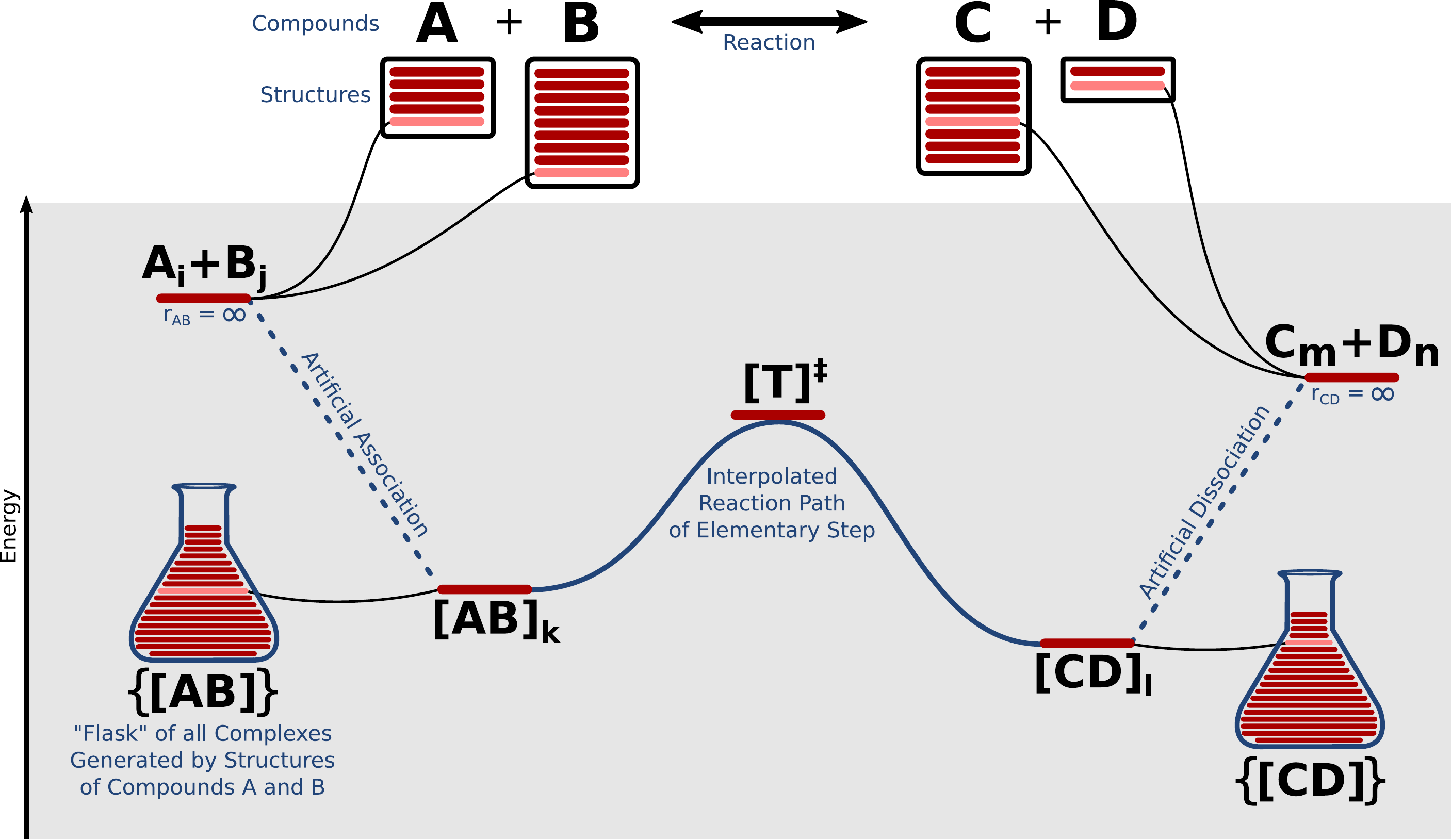}
 \end{center}
 \caption{\label{fig:flasks}\small Relationship of Compounds, Structures, Reactions, Elementary Steps, and different molecular complexes involved in an exploration of a single model reaction $A+B \rightarrow C+D$.}
\end{figure}

When exploring reactions or elementary steps, the key focus and challenge is usually to find the transition state(s) connecting the two sides of the reaction/elementary step.
Here, a conceptual problem arises from the description of the start and end point of such a reaction.
As shown in Figure~\ref{fig:flasks}, there are multiple possible end-point definitions in configuration space, where the reactant structures on both sides of the reaction arrow are
associated according to weak (typically dispersive) interactions that lead to local minima representing associated reactant molecules at zero temperature (\textit{i.e.}, at zero kinetic energy in a classical picture).
While the choice of end point does not affect statements pertaining to the existence of a given reaction, the kinetic analysis may be altered dramatically depending on the end point definitions.

One option is to separate the structures in each of the complexes on both sides of the transition state.
The energies of the separated structures are then evaluated (possibly after geometry optimization). 
These separate energies, $E(A_i)$, $E(B_i)$, $E(C_m)$ and $E(D_n)$,
where different structures of the compounds are labeled by lowercase letters, can then be combined and result in a first definition of the reaction energy and reaction barrier:
\begin{align}
    \Delta E^\ddagger_1 &= E([T]^\ddagger) - E(A_i) - E(B_i)  \label{eq:de3}\\
    \Delta E^\text{rxn}_1 &=  E(C_m) + E(D_n)- E(A_i) - E(B_i) \label{eq:ea3}
\end{align}
This choice is the de-facto standard for manual quantum chemical studies of elementary steps and reactions.

Starting at the transition state $[T]^\ddagger$ of our reaction $A+B \rightarrow C+D$, it is conceptually straightforward to follow minimum energy paths (MEPs) into local minima of both reactant valleys.
The three specific points, \textit{i.e.}, left-hand side complex $[AB]_k$, right-hand side complex $[CD]_l$, and transition state $[T]^\ddagger$), are typically defined on the same Born-Oppenheimer potential energy surface. Naturally, we assume that they
have been calculated within the same approximations (same basis sets, methods, solvent model and so forth).
Calculating a reaction barrier $\Delta E^\ddagger$ and overall reaction energy $\Delta E^\text{rxn}$ from these three points,
\begin{align}
    \Delta E^\ddagger_2 &= E([T]^\ddagger) - E([AB]_k) \label{eq:de1}\\
    \Delta E^\text{rxn}_2 &= E([CD]_l) - E([AB]_k) \label{eq:ea1}
\end{align}
is a second way toward rate constants and a kinetic model.

However, following the IRC scan delivers an energy on each side of the reaction arrow that corresponds to the first local minimum encountered.
Note that this may be an energetically highly unfavorable structure (\textit{e.g.}, a conformer that is not the lowest-energy conformer).
Hence, calculating activation energies of such a structure will underestimate the true reaction barrier and then overestimate the reaction rate. In other words, the activation energy to reach the high-energy local-minimum structure is not taken into account.
The resulting kinetic model will then not describe the reaction accurately.

A remedy for this problem is to correct the rates and barriers w.r.t. all other conformers or local minima of complexes. Whereas the sampling of conformers of a single \textit{structure} is available from its \textit{compound}, such a sampling needs to be generated for the latter case of complexes.
In Fig.~\ref{fig:flasks}, we call this set of non-bonded complexes the (virtual) flask\cite{Simm2017} of all complexes.
In practice, reaction rates can be modified with an additional probability of arriving at this end-point structure in the first place.
Hence, possible strategies are
(i) to replace the energy of the IRC end-point structure with the one from the flask structure or conformer in the set that features the lowest energy,
(ii) to extract the probability of the IRC end-point based on a Boltzmann distribution of all known conformers or flask structures for some temperature,
or (iii) to simulate the molecular dynamics in the flask and extract a probability for the IRC end-point.
Note that strategies (i) and (ii) assume that the barriers for conformational changes or rearrangement of the complexes in the flask are low (\textit{i.e.}, these options assume a pre-equilibrium among conformers or flask structures).
Applying any of the three strategies above will generate a third route toward kinetic modeling, one that corrects for the overall content of the flasks.

This last option, should give the most accurate description of the path of interest.
However, depending on the chosen strategy, the generation of the kinetic model is then the most involved one as it requires explicit sampling of structures in flasks on every side of an elementary step.
Even though the flask for the left-hand side can be reused for all other reactions of $A$ with $B$, the general sampling of all complexes of all pairs of compounds will be unfeasible
for automated explorations of larger networks.

By contrast, the first option is the one that is most easily evaluated.
The Eqs.~(\ref{eq:de3}) and (\ref{eq:ea3}) rely only on data that will already exist after the sets of all conformers of each of the compounds were generated.
In either way, the number of  conformer sets to sample only grows linearly with the number of compounds, while the number of sets of complexes grows at least quadratically (and even cubically for complexes of three compounds).

It is important to note that, especially when also considering an environment (\textit{e.g.}, a solvent in a microsolvation approach\cite{Bensberg2022}), the overall kinetics of an elementary step under consideration may be manipulated by the way a flask is set up.
For instance, for polar reagents it is easy to imagine, that in simple vacuum calculations the artificial association may be so strong that the overall barrier of the reaction may be calculated as negative according to Eq.~\ref{eq:ea3}.
For this reason, \textsc{Chemoton} in its current version provides a fitted spline of each computed energy path with all elementary steps.
This path leading to the complexes on either side of the transition state can be evaluated to calculate the energies as given in Eqs.~(\ref{eq:de1}) and (\ref{eq:ea1}).
Furthermore, the spline captures the shape of a path similar to the MEP (note, however, that deviations from the shape of the MEP may occur if IRC scans are accelerated by pseudo-Newton--Raphson algorithms).
Future releases will present advanced functionalities to construct flasks automatically from all end points of MEPs ending in the same set of complexes, which samples flasks without any extra computational work.

\section{Computational Methodology}
\label{sec:computational_methodology}
All data reported in this work were generated with the \textsc{SCINE} software framework\cite{Scine}.
\textsc{Chemoton}\cite{chemotonZenodo} or \textsc{Python3} scripts based on the \textsc{SCINE} \textsc{Python3} packages including \textsc{Chemoton} were run to steer data generation and to control the workflow. 
All data were stored in and processed from the \textsc{SCINE Database}\cite{databaseZenodo}.
All calculations were processed by \textsc{Puffin}\cite{puffinZenodo} instances, which internally interfaces \textsc{Readuct}\cite{Vaucher2018, Brunken2020}, \textsc{Molassembler}\cite{Sobez2020, Sobez2020a}, and the \textsc{SCINE Utilities}\cite{Bosia2020}.
For our work here, we employed pre-release versions of \textsc{Chemoton 2.0}\cite{chemotonZenodo}, \textsc{Puffin}\cite{puffinZenodo} and \textsc{SCINE Database}\cite{databaseZenodo}.

Autonomously launched electronic structure calculations were carried out either with \textsc{xTB}\cite{Bannwarth2020} (extended tight binding calculations, xTB) or SCINE \textsc{Sparrow}\cite{Husch2018, Bosia2020b} (density functional based tight binding, DFTB).
All extended tight binding calculations were carried out with the GFN2-xTB\cite{Bannwarth2019} model.
All density functional based tight binding calculations were carried out with the DFTB3\cite{Gaus2011} model employing the the parameters from Ref~\citenum{Gaus2013}.

All structures generated here will be made available on Zenodo upon publication of this work.

\section{Results}
\label{sec:results}
We have assembled a set of prototypical reactions. All reaction equations along with the reaction labels that will be used in the following are collected in Table~S1 in the Supporting Information.

For many of the reference reactants several different reactions, starting from the same set of reactants but leading to distinct sets of products, are included in our test set.
\textsc{Chemoton} explorations shall find not only one but (ideally) all reactions arising from given reactants. 
That is why we ran one exploration per reactant set (\textit{vide infra}), but for many cases expected \textsc{Chemoton} to recover more than one reference reaction.
To reflect this set-up, the reference reactions are labeled as \textit{"starting materials number"."reaction number"}.
Hence, in the following, if we, \textit{e.g.}, refer to reactions 7-9 we actually refer to all reactions expected from the reactant set 7, from set 8, and from set 9.

First, the test set includes the Zimmerman test set from Ref.~\citenum{Zimmerman2013} with the xyz-structures of reference reactants and products taken from Ref.~\citenum{Rasmussen2020} (reactions 1--27).
Reactions~28 are the unimolecular decomposition reactions of 3-hydroperoxypropanal as studied by Grambow \textit{et al.} in Ref.~\citenum{Grambow2018} with the xyz-structures taken from Ref.~\citenum{Rasmussen2021}.
Reactions 29 to 38 represent a slightly modified form of the test set for pericyclic reactions by Houk and coworkers \cite{Guner2003}.
Moreover, we added the small molecule test set of Lavigne \textit{et al.} \cite{Lavigne2020} (reactions 39--45).
With reactions 46--51 we include transition-metal-chemistry test cases from Ref.~\citenum{Zimmerman2015}. 
To achieve an even higher diversity regarding the types of transition metals, we included the TMB11 set (reactions 54--60).\cite{Chan2019}
Finally, we add reactions that we expect to be challenging---or even impossible---to uncover for our current algorithms (see also Section~\ref{sec:limitations} below):
HCN isomerization (reaction 61) is expected to be difficult due to the fact that the reaction coordinate, and hence, the artificial force applied in both Newton trajectory algorithms outlined in Section~\ref{sec:nt} is on the rotation axis of the linear molecule.
Reactions 62 and 63 are known to be quasi-barrierless.\cite{Quapp2007}
With reactions 64 to 68 from Refs.~\citenum{Goerigk2017} and~\citenum{Costentin2003} we include radical reactions. 
The \textit{cis-trans} isomerisation of azobenzene (reaction~69) cannot be represented in terms of connectivity changes.
Finally, reaction 70 is a bifurcation reaction.\cite{Yamamoto2011, Lee2020}

In this work, we focus on the question whether our algorithms are capable of finding certain new compounds. Hence, we removed all reference reactions that reproduce the reactants (\textit{i.e.}, where reactants and products only differ w.r.t.  their atom ordering) and only kept one of several elementary reactions resulting in the same compound.

Overall, our test set comprises $70$ distinct sets of reactants
and a total number of $184$ reference product sets
expected to arise from these reactants via the $184$ different reference reactions.
We are aware that this test set is of course not complete in any respect. Still, it covers a wide range of chemical reactions, hence, allowing us to assess the capacities of our software.

The calculations were carried out in a fully automated manner:
For each set of reactants a separate exploration was launched in one designated database. This means, for the 70 different reactant sets, 70 test explorations were launched in 70 databases. After loading the reference reactants and products into the database these are subjected to structure optimizations and are sorted into \textit{compounds} by \textsc{Chemoton}. The reference reactants are then activated for the exploration in such a way that elementary step trials are set up by the elementary step search engine. If a \textit{reaction} that connects the starting \textit{compounds} with the reference product \textit{compounds} is generated, the reference reaction will be considered found. If the structure optimization of any of the reference structures fails or results in dissociation of the molecule, the corresponding reference reaction will be counted as not found.

For the results that we present in the following, we limited ourselves to the exploration of one-step reactions, \textit{i.e.}, we did not explore the reactivity of the newly found intermediates, generating only a minimal reaction network for each of the reactant combinations.
Further, we opted to only consider bimolecular reactions between the given start structures for reactions with several reactants and only unimolecular reactions for those with one reactant.
Intermolecular self-reactions were disabled for all cases apart from reactions~1 and~36, where this is exactly the behavior of interest.

The results of the test explorations are, of course, heavily dependent on the number and type of elementary step trials that are carried out.
In \textsc{Chemoton}, when used without filters, there are a few key settings that control the number of reaction trials that are set up.
These options used for the generation of elementary step trials are summarized in Table~\ref{tab:estep_options} and will be further commented on during the discussion of the results.
For a more extensive list, including technical settings, we refer the reader to the Supporting Information.
No reactive site filters were applied, \textit{i.e.}, all reactive coordinates that are in agreement with the given options were considered.

\begin{table}[h!]
    \centering
    \caption{Overview of key \textsc{Chemoton} settings for the generation of elementary step trials with the NT1 and NT2 algorithms.
             (--- : not applicable, * : reactions: 27, 38, 47--49, 51, 54, 55, 57--60, 69.)}
    \label{tab:estep_options}
    \begin{tabular}{llcc}
     \hline \hline
     Molecularity & Setting & NT1 & NT2 \\
     \hline
     Bimol.  & Multiple attack directions per fragment & $\times$  & $\times$ \\
             & \# rotamers                        & 1 (none)  & 1 (none) \\
             & Max. \# intermol. bond formations       & $2$       & $2$      \\
             & Max. \# intramol. bond formations       & ---       & $0$      \\
             & Max. \# intramol. bond dissociations    & ---       & $0$      \\
             & Multiple conformers per compound        & $\times$  & $\times$ \\
     \hline
     Unimol. & Max. \# bond modifications              & ---       & $3$ $(2^*)$ \\
             & Max. \# bond formations                 &$2$ $(1^*)$&$2$ $(1^*)$ \\
             & Max. \# bond dissociations              & $1$       & $1$      \\
             & Multiple conformers per compound        & $\times$  & $\times$ \\
     \hline \hline
    \end{tabular}
\end{table}

The chosen settings attempt to find a maximum of two new intermolecular bonds to be formed during bimolecular reactions.
For the NT2 algorithm, a maximum of three simultaneous bond modifications was considered in unimolecular reactions with not more than two bond formations and one bond dissociation being explicitly considered at the same time.
The steep incline in reactive coordinates resulting from allowing two associations instead of one results in enormous numbers of calculations for some of the studied cases.
This is why, for these examples, we chose to reduce the maximum number of associations to one, if the expected reactions did not specifically require two concerted bond formations (see Table~\ref{tab:estep_options}).

The NT1 algorithm, as explained in Section~\ref{sec:nt}, does not allow for enforcing bonds to break and form simultaneously, which is why here \textit{either} a maximum of two concerted bond formations  \textit{or} one bond dissociation was tried. 

We note that this setup is far from a brute-force attempt aiming to find all elementary reaction steps.
The options shown in Table~\ref{tab:estep_options}, however, represent a reasonable approach to what currently constitutes a good balance between computational cost (with semi-empirical methods) and success rate without requiring much \textit{a priori} human knowledge of the expected chemistry.

For these settings with GFN2 as the electronic structure method, the explorations resulted in 118 (64~\%) and 145 (79~\%) reactions found out of the 184 reference reactions with the NT1 and NT2 algorithms, respectively. These and other selected key statistics of these runs are summarized in Table~\ref{tab:statistics_all}.

\begin{table}[htbp]
    \centering
    \caption{Share of recovered reference reactions and total number of reactions found during runs with different algorithms and electronic structure methods. The number of elementary step trial calculations and the percentage of these that resulted in elementary steps are listed in the third last and last columns, respectively. 
    The percentage of reactions found with DFTB3 is given with respect to the subgroup of reference reactions, for which DFTB3 parameters were available.}
    \label{tab:statistics_all}
    \begin{tabular}{@{\extracolsep{4pt}}ccrrrrr}
    \hline \hline
                           &           & \multicolumn{2}{c}{Reactions} & \multicolumn{3}{c}{Elementary Steps} \\ \cline{3-4} \cline{5-7}
    Method                 & Algorithm & Reference     & Overall      & Trials & Found & Success Rate \\ 
                           &           & Found in \%   & Found        &        &       & in \%  \\ \hline
    \multirow{ 2}{*}{GFN2} & NT1 & 64 & 3,329 & 303,167 & 31,424 & 10.4 \\ 
                           & NT2 & 79 & 8,368 & 3,746,424 & 381,857 & 10.2 \\ 
    DFTB3                  & NT2 & 69 & 8,849   & 2,291,202 & 236,091 & 10.3 \\ 
    \hline \hline
    \end{tabular}
\end{table}
For a detailed overview about which reference reactions were found and missed see Table~S3 in the Supporting Information. The information from Table~\ref{tab:statistics_all} separated for each test case is available in Table~S4 in the SI.

The fact that, as outlined in section~\ref{sssec:NT1}, the NT1 algorithm only requires two reactive fragments to be specified as input, compared to the combination of reactive pairs in the NT2 algorithm, results in a lower number of elementary step trial calculations.
However, here, due to its overall higher detection rate, if not stated otherwise, we will focus on the analysis of the NT2 results in the following.

One might argue that even the 21\,\% of reactions that were not accessible to the generally more successful NT2 algorithm are unsatisfactory.
However, in the following we will explain why a 100\,\% detection rate was never to be expected and, more importantly, how the current rate can be improved systematically.

As outlined above, we deliberately assembled a challenging test set of reactions. 
To begin with, we included representatives of reactions that go beyond the known limitations of our current infrastructure in order to expose Chemoton to a serious test.
These limitations, \textit{e.g.}, homolytic bond cleavages, will be discussed in Section~\ref{sec:limitations} below, where we also address how they may be overcome.

\begin{figure}[htbp]
 \begin{center}
  \includegraphics[scale=.70]{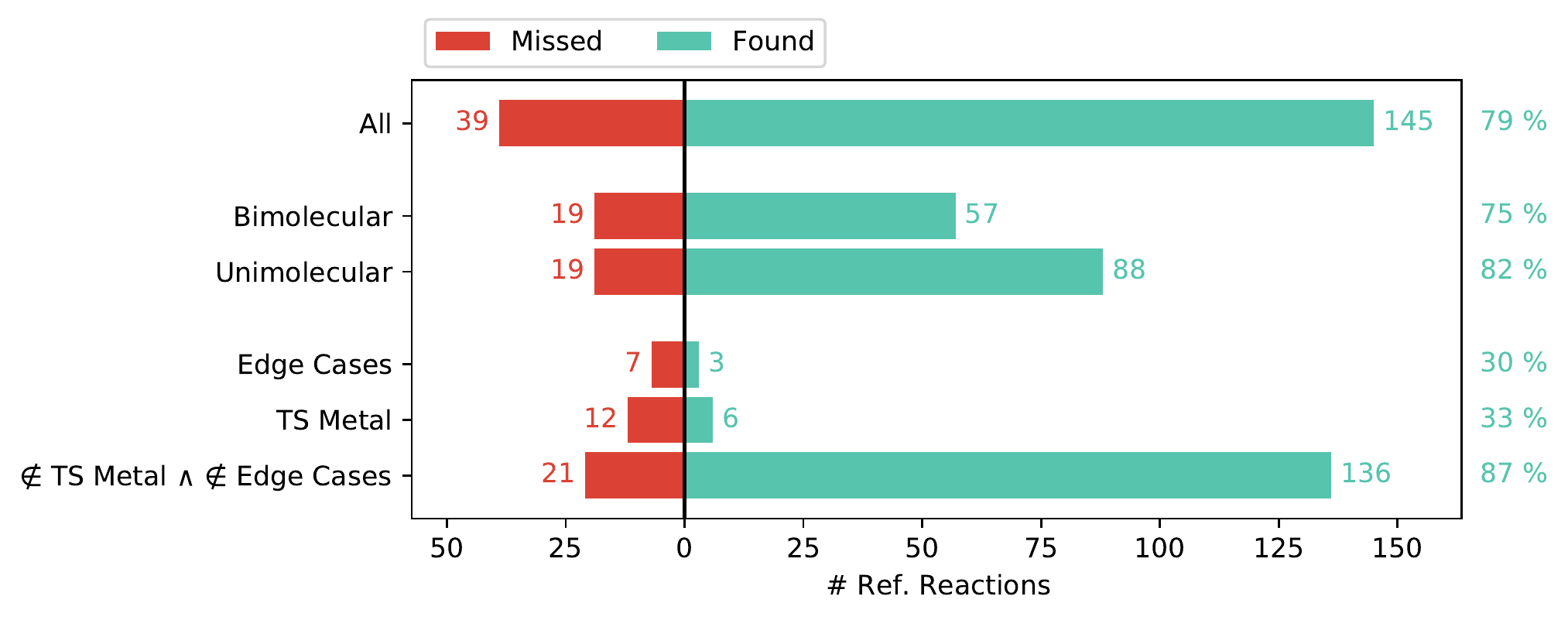}
 \end{center}
 \caption{\label{fig:categorized_success}\small Number of reference reactions found (green) and missed (red) among different subcategories of the test set of reactions. The data shown are generated with the NT2 algorithm and the GFN2-xTB method. The category ``edge cases'' includes reactions 53 (which is trimolecular), 61 (HCN isomerisation), 62, 63 (which are quasi-barrierless), 64 (a radical recombination), 66 (which is supposed to result in a triplet state), 68 (a homolytic dissociation), 69 (the \textit{cis-trans} isomerisation of azobenzene) and 70 (a bifurcation reaction).}
\end{figure}

As can be seen in Figure~\ref{fig:categorized_success}, the detection rate among these edge cases is  low, as expected, with only 3 out of 10 reference reactions being found.

Another striking feature that becomes apparent from Figure~\ref{fig:categorized_success}, is that with only 33\,\%, the share of found reference reactions is particularly low among those that involve transition metal complexes.
As these types of compounds are known to be challenging for electronic structure methods (and in the reference literature were analyzed with more reliable electronic structure models), the tight-binding electronic structure models will face problems here as we shall show below.
We also note that the unambiguous perception of molecular graphs based on bond orders is a critical component and particularly challenging for the complex bonding situations in transition-metal compounds.

The layered design of \textsc{Chemoton} described in Section~\ref{sec:theory} allowed us to repeat our analysis with another electronic structure model. Already DFTB3 provides us with a first impression of the effect of the approximate electronic structure method on the exploration outcome.
DFTB3 resulted in 100 (69\,\%) out of the 144 reference reactions for which DFTB3 parameters were available to be found. For these reactions the NT2-GFN2 success rate was 84\,\%. 
25 of the reference reactions detected with GFN2 were not recovered with DFTB3 and 4 with DFTB3, but not with GFN2.

These results clearly demonstrate that the chosen electronic structure method affects what reference reactions are found and
hint at the fact, that both false positives and false negatives are to be expected when applying a fast, but not very accurate electronic structure model.
Obviously, more accurate, but also computationally more expensive models such as generalized gradient approximation DFT with efficient density fitting need to be employed. For the whole procedure to remain computationally feasible, the semi-empirical tight-binding models may be used for exploratory purposes only. Subsequent DFT calculation can be used to refine (i) stationary structures found, (ii) successful elementary step trials, or (iii) to launch completely new elementary step trials.

Exclusion of the aforementioned edge cases and of all transition-metal reactions, leaves us with a test set of 157 reactions, of which 136, \textit{i.e.}, 87\,\% were successfully recovered.
In the following, we briefly outline which options we recommend to change if the success rate w.r.t. the reference data shall be increased even further, without changing the electronic structure method.

\begin{figure}[htbp]
 \begin{center}
  \includegraphics[scale=.90]{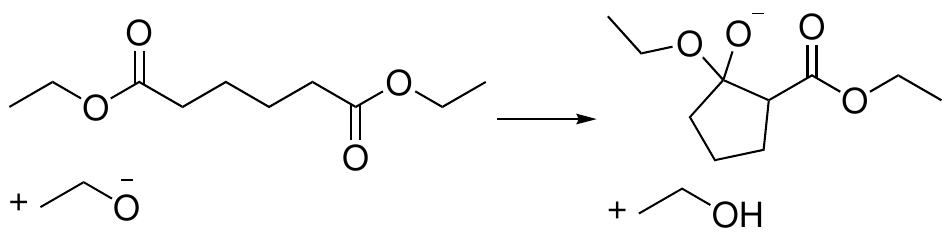}
 \end{center}
 \caption{\label{fig:rct_42}\small Reaction 42 of the test set.}
\end{figure}

One major simplification that we opted for in this work is the neglect of conformational variety. For each compound one arbitrarily chosen conformer was used as the reactant. However, our software infrastructure also  provides the option to generate conformers and to consider all of them as reactive. 
While this results in an increase of computational effort, which scales with the number of conformers of the reactant for unimolecular reactions and the product of the number conformers of both reactants for bimolecular reactions, it may still be of crucial importance for some reactions.
Consider as an example reaction 42, where a long alkyl chain undergoes ring closure as depicted in Figure~\ref{fig:rct_42}.
This example was taken from the work by Lavigne \textit{et al.}~\cite{Lavigne2020} that exploits the conformational exploration of activated structures.

\begin{figure}[htbp]
 \begin{center}
  \includegraphics[scale=1.0]{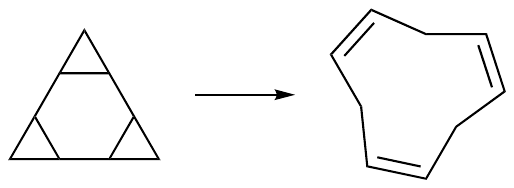}
 \end{center}
 \caption{\label{fig:rct_37}\small Reaction 37 of the test set.}
\end{figure}

\begin{figure}[htbp]
 \begin{center}
  \includegraphics[scale=1.0]{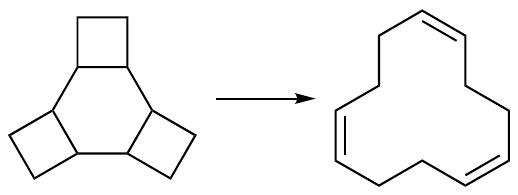}
 \end{center}
 \caption{\label{fig:rct_38}\small Reaction 38 of the test set.}
\end{figure}

As can be understood from Figure~\ref{fig:categorized_success}, bimolecular reference reactions were less likely to be found than unimolecular ones with detection rates of 75\,\% and 82\,\%, respectively. 
The number of rotamers and, especially, the number of attack directions per reactive fragment are two options that only apply to bimolecular reactions and both were set to one (see Table~\ref{tab:estep_options}), \textit{i.e.}, to the most minimalistic option.
Furthermore, especially for bimolecular but also for unimolecular reactions, as outlined above, the type of trial reactive coordinates was limited within narrow bounds. As we explained in Section~\ref{sec:nt}, choosing certain bonds to be formed or broken does not limit the scan to the modification of \textit{only} these bonds. For example, the cycloreversion reactions in reactions 37 and 38 shown in Figures~\ref{fig:rct_37} and~\ref{fig:rct_38} were both reached, even though the three concerted bond dissociations occurring there, were not enforced explicitly.

Still, the choice of coordinates has a critical effect on the computational effort and on the detected elementary steps, as we will outline for the example of the unimolecular reactions of reactant 28, 3-hydroperoxypropanal.
With 54 reactions we have a vast amount of reference data to recover for this case, making it a perfect candidate for a more detailed analysis.
In Table~\ref{tab:mod_statistics_28}, we show the results depending on the number of bond formation and dissociation reactions explicitly included in the trial reactive coordinate for reaction set number 28. 

\begin{table}[htbp]
    \centering
    \caption{The number of elementary step trial calculations carried out for reactant combination~28 resolved in terms of the number of bond formation (form.) and dissociation reactions (break) explicitly included in the reactive coordinate. The success rate in the last column is the share of calculations that resulted in an elementary step. Note that the elementary steps are not deduplicated, \textit{i.e.}, several can represent the same chemical conversion.}
    \label{tab:mod_statistics_28}
    \begin{tabular}{ccrc}
    \hline \hline
    \# form. & \# break & \# Elementary Step Trials & Success Rate/\% \\ \hline
              0 &           1 &         11 & 18         \\
              1 &           0 &         55 & 15         \\
              1 &           1 &        605 & 22         \\
              2 &           0 &       1,485 & 30         \\
              2 &           1 &      16,335 & 23         \\
    \hline
     0--2 &  0--1 &      18,491 & 24         \\
    \hline \hline
    \end{tabular}
\end{table}

With 16,335 out of 18,491 elementary step trial calculations (88\,\%), the trial coordinates composed of two bond formations and one dissociation largely dominate, as to be expected based on combinatorial arguments. 
The success rate, \textit{i.e.}, the share of trials that results into the generation of an elementary step, for these trials is 23\,\%.
This number indicates that neglecting these coordinates cannot be generally recommended. Inclusion of further reactive coordinates is likely to yield even more elementary steps and, possibly, reactions.

\begin{figure}[htbp]
 \begin{center}
  \includegraphics[scale=0.9]{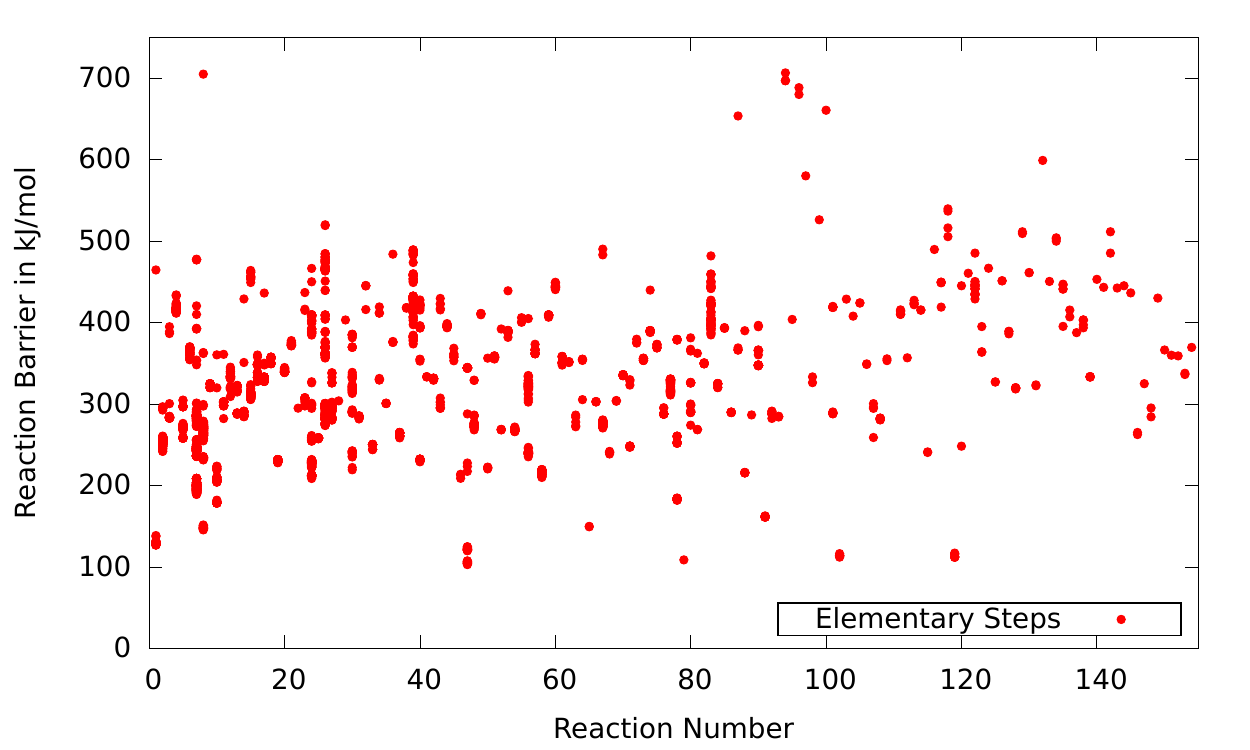}
 \end{center}
 \caption{\label{fig:barrier_distributions}\small Reaction barriers for all elementary steps explored with the NT2 algorithm and GFN2-xTB for reaction 28.
Elementary steps are stacked according to the reaction they belong to.}
\end{figure}

Out of the 54 reference products 47 were found successfully.
However, it is important to stress that beyond these hits and misses that we expected explicitly based on the reference data, \textsc{Chemoton} did succeed in finding a plethora of other elementary steps and reactions: The 4,365 (not deduplicated) elementary steps belong to 154 reactions.
In principle, it is possible that many elementary steps simply have very high barriers and are thus only relevant in special cases. However, as
shown in Figure~\ref{fig:barrier_distributions}, this is not the case for reaction set number 28.

Many of the elementary steps that were found beyond the expected results display barriers that are close in energy to the barriers of expected reactions. Hence, Chemoton was able to provide a more detailed picture of the accessible reaction space than what was known from the literature.
This finding also holds true for other reaction sets in this test.

\subsection{Current Limitations}\label{sec:limitations}
In its present version of \textsc{Chemoton} has known limitations, which is not surprising considering the breadth and depth of tasks to be carried out autonomously.
Given the flexibility of the software described here, these limitations will be alleviated in future releases. However, for sake of completeness and also to guide production explorations by others, we dwell on some of the short-comings.

A challenge in explorations is tracking and assigning spin states.
Many of the successive steps described in the previous sections offer the option to change or reassign the spin state of a structure under consideration.
Imagine the combination of two doublet structures into one reactive complex, which may result in a structure to be considered as a spin triplet or singlet, which will typically require both options to be tracked in order to make sure that the proper ground state is chosen. Apart from the fact that this will face the nagging issue of the reliability of spin-state energetics\cite{Reiher2001,Salomon2002,Reiher2002,Harvey2004,Swart2016}
it may require specific techniques to enforce certain spin distributions (such as constrained DFT settings\cite{Kaduk2012})
in the somewhat artificial reactive-complex structures from which searches for transition states and reaction coordinates are started.
Even worse, the optimal spin state may change during the course of a single step due to a two-state-reactivity situation\cite{Schroder2000},
for which one is then advised to monitor close-lying states in a molecular-propensity calculation\cite{Vaucher2016b}
with ultrafast quantum chemical methods on the fly. Although such schemes can be integrated into the exploration
algorithm, they have not yet become available in the present version.
Currently, it is only possible to assign a spin state to the exploration procedure that will be applied for all steps and possible combinations of spin will be resolved by a single fixed rule.

The reaction finding algorithms can be improved on. For instance, no explicit exploration and handling of bifurcations in reaction paths is currently implemented. 
This limitation is probed with reaction 70.1 and 70.2 which are bifurcations of the same transition state.
At best, our current results are statistically distributed across possible bifurcated paths based on the numerics of the specific step trials.
For this particular issue, possible solutions have already been reported in the literature (see Refs. \citenum{Lee2020, Ito2020}) that will be tested and adapted in future work.

In a few special cases, it is required to artificially break the point-group or local 
symmetry of reactants in order to facilitate the correct reaction.
This is mostly a remedy for the technical issue of optimizations converging to saddle points instead of minima.
Another problem case is highlighted in reaction 61 in the test set: the isomerization of HCN requires breaking of the $C_{\infty\nu}$ symmetry 
in order to move the hydrogen atom around the carbon and nitrogen atoms.
A preliminary setting to enable these types of symmetry breaking exists in \textsc{Chemoton}, however, it needs to be calibrated so that it does not skew the actual exploration, but only breaks the symmetry of a given reactant in a minimal way.

Elementary steps featuring reaction coordinates that cannot be well-described in terms of atoms being pushed/pulled onto each other (\textit{e.g.}, the \textit{cis}-/\textit{trans} isomerization of azobenzene) are not easily explorable.
While these reactions may accidentally be found even with the current version, a targeted search for such reactions will be needed.

Furthermore, the current implementation of the step trials expect a transition state to exist for each elementary step.
Many cases in which no reasonable transition state can be found may actually be artifacts that are related to simple physical association of reagents (as in the physisorption process on surfaces); see the discussion in Section~\ref{sec:flask}.
For all other barrier-free transitions, a special book keeping procedure must be introduced so that generated networks are complete and explore regions that are gated behind them.
Furthermore, these transitions require logging so that accurate kinetic modeling based on \textit{e.g.} diffusion mechanisms may be based to them.
Example cases for such transitions are ion--counter-ion associations that form tight ion pairs or salts.

Finally, elementary reactions with the same reactant and product structures are currently not tracked.
This is a result of the chosen definition of structures that are labeled ``the same''.
Our definition implies that single atoms of the same element are indistinguishable.
Therefore, \textit{e.g.}, swapping only two hydrogen atoms in a reaction does not constitute an elementary step.
This problem can be addressed with exact atom index mapping along the path. However, it does somewhat model experimental circumstances where isotope labeling would be required to observe these reactions.

\section{Conclusions}
\label{sec:conclusion}
Here, we presented the modular software framework Chemoton 2.0 that enables the automated exploration of reaction networks of any chemical reactive system.
We described the software architecture and introduced key algorithms.
We demonstrated how our framework can technically be extended by means of predefined interfaces. 
All software modules described in this work will be released open source and free of charge;
for more information about the software we refer to our web page\cite{Scine} and
the corresponding \textsc{GitHub} page\cite{ScineGitHub}.

To demonstrate the capabilities of the current feature set, we assembled and explored a chemically diverse set of reactions. This data set is made publicly available in order to provide a suitable starting point for future benchmarking studies of mechanism-exploration algorithms.
We showed that \textsc{Chemoton} found many molecular transformations that may not have been noticed in a traditional manual  exploration.
Already with basic settings for finding reactions based on elementary step trials we demonstrated that \textsc{Chemoton} had a success rate of $79\,\%$ for finding reactions known from the literature.
Beyond these, \textsc{Chemoton} reported 44 times more reactions than what was documented in the
literature, many of which with low barriers.

It is important to note that the failure rate of $21\,\%$ seems to be high, but is in fact
due to drawbacks of the electronic structure model applied and due to features missing in the current version of 
\textsc{Chemoton} (\textit{e.g.}, for finding reactions with bifurcation), which will be improved on in future releases of the software.

With the general framework of \textsc{Chemoton} 2.0 it is now possible to map out extensive
reaction networks of reactive systems containing elements from across the periodic table. While this capability makes our exploration system well-suited for the elucidation of kinds of chemistries, it may also be exploited to advance exploration algorithms. For instance, the
networks obtained in a brute-force fashion so far can now be taken as a data reservoir to assess the predictive power of chemical reactivity
concepts\cite{Grimmel2021}, which in turn can then be exploited to filter 
elementary step trials based on first-principles heuristics.\cite{Bergeler2015, Grimmel2019}
Clearly, various directions for future development and extensions are obvious (see also the previous section); we will consider interleaving explorations with \textsc{Chemoton} with microkinetic modeling
approaches such as KiNetX \cite{Proppe2016,Proppe2019} and extending the search algorithms for elementary-step trials towards molecular dynamics approaches with tailored biasing schemes\cite{Laio2002,Barducci2011,Wang2014,Valsson2016,Shannon2018,Grimme2019,Chen2022}.

\section*{Supporting Information}
Reaction equations of the reference reactions, additional information about technical settings employed in the explorations, and a more detailed breakdown of the results are provided as additional material in the Supporting Information.

\section*{Acknowledgments}
\label{sec:acknowledgments}
This work was financially supported by the Deutsche Forschungsgemeinschaft (DFG) (GZ: UN 417/1-1) and by the Schweizerischer Nationalfonds (SNF) (Project 200021\_182400).

\providecommand{\latin}[1]{#1}
\makeatletter
\providecommand{\doi}
  {\begingroup\let\do\@makeother\dospecials
  \catcode`\{=1 \catcode`\}=2 \doi@aux}
\providecommand{\doi@aux}[1]{\endgroup\texttt{#1}}
\makeatother
\providecommand*\mcitethebibliography{\thebibliography}
\csname @ifundefined\endcsname{endmcitethebibliography}
  {\let\endmcitethebibliography\endthebibliography}{}

\end{document}